\documentclass[a4paper,onecolumn,11pt,accepted=2019-10-09]{quantumarticle}
\pdfoutput=1
\usepackage[utf8]{inputenc}
\usepackage[english]{babel}
\usepackage[T1]{fontenc}
\usepackage{amssymb}
\usepackage{amsmath}
\usepackage{hyperref}
\usepackage[numbers,compress]{natbib}
\usepackage{empheq}
\usepackage{bm}

\usepackage{tikz}
\usepackage{lipsum}

\newcommand{\ann}{\hat{a}}
\newcommand{\adg}{\hat{a}\dg}
\newcommand{\beq}{\begin{equation}} 
\newcommand{\eeq}{\end{equation}}
\newcommand{\bqa}{\begin{eqnarray}} 
\newcommand{\eqa}{\end{eqnarray}}
\newcommand{\nn}{\nonumber}

\newcommand{\dg}{^\dagger}
\newcommand{\rt}[1]{\sqrt{#1}\,}

\newcommand{\bra}[1]{\langle{#1}|} 
\newcommand{\ket}[1]{|{#1}\rangle}
\newcommand{\an}[1]{\langle{#1}\rangle}
\newcommand{\ban}[1]{\big\langle{#1}\big\rangle}

\newcommand{\op}[2]{\left|{#1}\rangle \langle{#2}\right|}

\newcommand{\fbo}[4]{\left[\begin{array}{c} {#1} \\ {#2} \\ {#3} \\ {#4} \end{array}\right]}

\newcommand{\ito}{It\^o} 

\newcommand{\sch}{Schr\"odinger} 
\newcommand{\hei}{Heisenberg}

\newcommand{\fp}{Fokker--Planck}

\newcommand{\up}{\uparrow}
\newcommand{\dn}{\downarrow}

\newcommand{\kup}{\kappa_\up}
\newcommand{\kdn}{\kappa_\dn}

\newcommand{\ddt}{\frac{d}{dt}}

\newcommand{\kminus}{\kappa_-}

\newcommand{\E}{{\rm E}}

\begin{document}

\title{Phase diffusion and the small-noise approximation in linear amplifiers: Limitations and beyond}

\author{Andy Chia}
\affiliation{Centre for Quantum Technologies, National University of Singapore, 3 Science Drive 2, Singapore 117543}
\email{photonicboy@gmail.com}
\author{Michal Hajdu\v{s}ek}
\orcid{0000-0002-8319-9566}
\affiliation{Centre for Quantum Technologies, National University of Singapore, 3 Science Drive 2, Singapore 117543}
\thanks{Current address is Keio University Shonan Fujisawa Campus, 5322 Endo, Fujisawa, Kanagawa 252-0882, Japan}
\author{Rosario Fazio}
\affiliation{ICTP, Strada Costiera 11, I-34151 Trieste, Italy}
\affiliation{Dipartemento di Fisica, Universit\`{a} di Napoli "Federico II", Monte S. Angelo, I-80126, Italy}
\orcid{0000-0002-7793-179X}
\author{Leong-Chuan Kwek}
\affiliation{Centre for Quantum Technologies, National University of Singapore, 3 Science Drive 2, Singapore 117543}
\affiliation{MajuLab, CNRS-UNS-NUS-NTU International Joint Research Unit, UMI 3654, Singapore}
\affiliation{National Institute of Education, Nanyang Technological University, 1 Nanyang Walk, Singapore 637616}
\orcid{0000-0002-0879-0591}
\author{Vlatko Vedral}
\affiliation{Centre for Quantum Technologies, National University of Singapore, 3 Science Drive 2, Singapore 117543}
\affiliation{Department of Physics, University of Oxford, Parks Road, Oxford, OX1 3PU, UK}
\orcid{0000-0003-4561-5124}
\maketitle

\begin{abstract}
  The phase of an optical field inside a linear amplifier is widely known to diffuse with a diffusion coefficient that is inversely proportional to the photon number. The same process occurs in lasers which limits its intrinsic linewidth and makes the phase uncertainty difficult to calculate. The most commonly used simplification is to assume a narrow photon-number distribution for the optical field (which we call the small-noise approximation). For coherent light, this condition is determined by the average photon number. The small-noise approximation relies on (i) the input to have a good signal-to-noise ratio, and (ii) that such a signal-to-noise ratio can be maintained throughout the amplification process. Here we ask: For a coherent input, how many photons must be present in the input to a quantum linear amplifier for the phase noise at the output to be amenable to a small-noise analysis? We address these questions by showing how the phase uncertainty can be obtained without recourse to the small-noise approximation. It is shown that for an ideal linear amplifier (i.e.~an amplifier most favourable to the small-noise approximation), the small-noise approximation breaks down with only a few photons on average. Interestingly, when the input strength is increased to tens of photons, the small-noise approximation can be seen to perform much better and the process of phase diffusion permits a small-noise analysis. This demarcates the limit of the small-noise assumption in linear amplifiers as such an assumption is less true for a nonideal amplifier. 
\end{abstract}

\section{Introduction}
\label{Introduction}

    It was shown in the 1960s that linear amplification of light is an inherently noisy process \cite{HM62,Hef62,Cav82}. As a result, the input phase suffers from an increased uncertainty during the course of amplification. The process by which this occurs in a phase-insensitive linear amplifier is phase diffusion, the same process responsible for the natural linewidth of single-mode lasers \cite{ST58,Lax67,Hen82,SSL93,SZ97}. Aside from fundamental interests, phase diffusion has also been shown to be useful for quantum random number generation \cite{JCS+11,XQM+12,AAJ+14,SVG19}. More recently it has also been used to demonstrate the applicability of phase squeezing as a means to reduce phase noise in linear amplifiers \cite{CSOCP19}.
    
    Phase diffusion in linear amplifiers has been studied with a variety of methods \cite{LS84,BSP89,Lu90,Ban91,TGKG92,VP94}. In general, one would like to characterise phase diffusion by calculating the diffusion coefficient, or by calculating a sensible measure of phase uncertainty (often the phase variance but not exclusively \cite{Ban91,Ban92,BB93,Hal93}). However, phase diffusion is inversely proportional to the photon number which makes the diffusion coefficient or variance difficult to calculate without approximations. The most widely used approximation assumes the input light to be sufficiently intense such that its photon number has an average much greater than its standard deviation (see e.g.~\cite{BSP89,Lu90}). The simplification gained from such an assumption is that one may replace the photon number by its average. Although this approximation is extremely useful and continues to be a standard method of analysis in quantum optics, it restricts one to field states with a narrow photon-number distribution. For example, weak coherent states, which may be produced by attenuating the output of a laser may not satisfy the small-noise approximation. These states have been shown to be useful for quantum cryptography \cite{HIGM95,ZDN15,GTZ02}. In addition, states with low number of photons are useful in numerous fields outside of quantum information science, such as single-molecule spectroscopy, light detection, and many others \cite{EFMP11,WWC+17,MGR+18}. We note that the small-noise approximation is, in spirit, similar to what others might call a ``semiclassical'' or ``mean-field'' approximation in quantum optics.\footnote{The semiclassical approximation replaces the bosonic annihilation operator $\ann$ by $\an{\ann}+\delta\ann$ and neglects $(\delta\ann)^2$ and terms with higher order in $\delta\ann$, where $\delta\ann$ is the deviation of $\ann$ from its average value.}
    
    A primary function of linear amplifiers is to boost the strength of a weak signal to detectable levels \cite{CDG+10}. For quantum linear amplifiers, increasing the signal strength comes at a cost---the amplifier must add noise if it is to operate according to the principles of quantum mechanics. The quantum limit on how much noise such an amplifier must add were investigated in the early days of linear amplifiers \cite{HM62,Hef62,Cav82}, and has been extended recently \cite{CCJP12}. If the input to the amplifier is a noisy signal (which is typically the case) this noise also gets amplified so in general the output consists of the amplified input signal, the amplified input noise, and the noise the amplifier adds. As a result, the signal-to-noise ratio degrades after amplification. This suggests that the small-noise approximation tends to get worse over the course of amplification, with the breakdown becoming more accute for weaker inputs. One can then try to optimise the signal-to-noise ratio by using an ideal linear amplifier---an amplifier which adds the least amount of noise required by quantum mechanics. It is therefore particularly interesting to know how well the small-noise approximation performs for an ideal linear amplifier as this is the most favourable scenario for the approximation to hold. In this paper we show that even for an ideal linear amplifier, the small-noise approximation is still inadequate in the regime of low-intensity inputs (few photons on average) and large amplifier gain. This is our main result which essentially demarcates the limit of the small-noise approximation in treating phase diffusion in linear amplifiers.
    
    An important aspect of the small-noise approximation is that it allows one to calculate the phase noise analytically. Naturally, one wonders if it is still possible to make some analytical progress when the small-noise approximation fails, or do we have to resort to numerics? In this paper we show that it is possible to generalise the small-noise approximation and obtain the phase uncertainty for weak inputs. In this treatment the phase uncertainty is given by a so-called inverse-number expansion. The expansion is checked against numerics and is also valid for nonideal linear amplifiers. We also derive a closed-form expression for the output phase noise under the small-noise approximation, which is then compared with the inverse-number expansion. An appreciable difference between the phase noise obtained from the small-noise approximation \text{\color{black} and the inverse-number expansion} can be seen for low-intensity coherent inputs where only a few photons are present on average (and more generally intensities that are on the order of unity). However, the applicability of the small-noise approximation is seen to be restored by including, on average, an additional ten photons or slightly less. This highlights the sensitivity of the small-noise approximation to the discreteness (or photon nature) of weak coherent inputs.
    
    The paper is organised as follows. We first introduce our linear-amplifier model in Sec.~\ref{Problem} (in terms of a master equation), and define our model of phase diffusion. Here we also review some essential features of the standard linear amplifier and state our measure of phase uncertainty. Here we also motivate the relevant parameter regime to operate the amplifier in and the importance of the coherent state as a simple model of the amplifier input. The small-noise approximation is then explained in Sec.~\ref{SmallNoise+InvNumExp} and we shall see how this approximation leads to a closed-form expression for the phase uncertainty. The details are left to Appendix~\ref{SmallNoiseVar}. The inverse-number expansion---which is our proposed method for going beyond the small-noise approximation---then follows quite naturally from this. The idea behind the inverse-number expansion can be simply explained so we leave its derivation and checks to Appendices~\ref{Result} and \ref{Validity}. Using the inverse-number expansion we then present our main result in Sec.~\ref{PhaseUncert}, where the phase uncertainty obtained by the two approaches are compared against each other. Here we shall see that a treatment of the phase noise beyond the small-noise approximation is necessary for a coherent input whose photon number is on the order of unity. This is despite the fact that such an input gets added as little noise as possible (a condition for which the small-noise approximation is least likely to breakdown during amplification). Of course, our chosen path to this result is not without any drawbacks so both the advantages and disadvantages of our approach are discussed in Appendix~\ref{PhaseWithinP}. We then conclude in Sec.~\ref{Conclusion} with a summary of our work and a discussion of the road ahead for extending the results of this paper.

\section{Phase diffusion in linear amplifiers}
\label{Problem}

    \subsection{Linear-amplifier model}
    \label{Model}

        There are a number of linear-amplifier models that one can choose from. A universal model in fact exists (a two-mode squeezer \cite{CCJP12}) provided the noise introduced by the amplifier is additive \cite{CHFKV17,CHF+19}. However, for phase diffusion, the focus of attention has been on a master-equation realisation of the linear amplifier and we will follow suit. An important element of the master-equation model is that it can be recast as a \fp\ equation---the quintessential equation for diffusive processes---which justifies calling the phase evolution (of the signal being amplified) phase diffusion. The master equation for a linear amplifier is given by
        \begin{align}
        \label{DensityOpEOM}
	        \ddt \rho(t) = \kdn \bigg[ \hat{a} \rho(t) \hat{a}\dg - \frac{1}{2} \hat{a}\dg \hat{a} \rho(t) - \frac{1}{2} \rho(t) \hat{a}\dg \hat{a} \bigg] 
	        + \kup \bigg[ \hat{a}\dg \rho(t) \hat{a} - \frac{1}{2} \hat{a} \hat{a}\dg \rho(t) - \frac{1}{2} \rho(t) \hat{a} \hat{a}\dg \bigg],
        \end{align}
        where $\kdn$ and $\kup$ are positive real numbers, and $\ann$ and $\adg$ are annihilation and creation operators satisfying the canonical commutation relation $[\ann,\adg]=\hat{1}$. Such a master equation may be derived by considering the interaction of a bosonic mode with a collection of two-level atoms. The terms proportional to $\kdn$ and $\kup$ collectively describe photon loss and gain respectively. Hence the coefficients $\kdn$ and $\kup$ can be thought of as rates at which photons are lost or gained. An input bosonic field prepared in the state $\rho(0)$ then experiences amplification when there is net gain i.e.~$\kup>\kdn$, a condition usually achieved via population inversion in the two-level atoms. The simplest way to see this is a direct calculation of the output photon-number using \eqref{DensityOpEOM},
        \begin{align}
        \label{<N(t)>}
	        \ban{\adg\ann}_t = G_t \, \ban{\adg\ann}_0 + \frac{\kup}{\kminus} \; \big( G_t - 1 \big)  \;,
	    \end{align}
        where we have defined the photon-number gain 
        \begin{align}
        \label{NumberGain}
	        G_t = e^{\kminus\,t} \;,  \quad  \kminus = \kup - \kdn  \;.
        \end{align}
        The output can be seen to comprise one part which is simply given by the input but now amplified by a factor $G_t$, and another portion, independent of the input, interpreted as noise due to spontaneous emission. Note that a given value of photon-number gain $G_t$ is determined by the product of the amplification time $t$ and the amplification rate $\kminus$. When $\kdn=0$, the model is said to describe an ideal linear amplifier, so called because in this case the amplifier adds the least amount of noise consistent with quantum mechanics \cite{Aga13}. From \eqref{<N(t)>}, the noise can be seen to be minimised when $\kdn=0$.
    
    \subsection{Amplifier parameter regime and input signal model}
    
        As we mentioned above, a linear amplifier serves primarily to increase a weak signal to measurable strengths \cite{Cav82,CDG+10}. The high-gain limit of linear amplifiers ($G_t\longrightarrow\infty$) is thus of special interest in research, both theoretically (such as addressing the amount of noise an amplifier must add \cite{Cav82,CDG+10}), and experimentally (such as the existing experimental efforts to build linear amplifiers in circuit QED \cite{CDG+10,LVH+14,RD18}). We therefore focus on the high-gain limit here as well since this is the most relevant regime.
        
        Aside from a suitable amplifier model we also need a suitable model for the amplifier input. For this we consider coherent states. Coherent states provide a particularly good model for the input signal because their signal-to-noise ratio is easily tunable via its average photon number. Another reason to consider coherent states is that quantum optical experiments are typically performed with lasers and the state of laser light can be effectively modelled as coherent. In fact, a recent experiment demonstrating phase diffusion and how it can be countered using squeezing was performed with a coherent state input \cite{CSOCP19}. Aside from these advantages of coherent states, there are also reasons for not considering other commonly used quantum-optical states: First, for phase diffusion, states with perfect angular symmetry---the vacuum, thermal, or number states---cannot be used as input models as they have a maximally diffused phase. Second, for the phase-preserving amplifier model of \eqref{DensityOpEOM}, one would not expect to observe the effects of nonclassical features on phase diffusion such as squeezing, sub-Poissonian statistics, or macroscopic superpositions (corresponding to the squeezed, number, and \sch-cat states). This is because such features will be completely lost even at moderate gains ($G_t=2$) \cite{Aga13,MW95,FM83,HFM85,MMS02}. It has also been shown that some nonclassicality can persist \cite{MG67,HZZ96,SVDASDM09,NMC10}. Despite this, one should keep in mind that one usually cares about nonclassical states because they are useful for quantum information processing (e.g.~see Refs.~\cite{JR07,GNM+04}). In these cases, most nonclassical states are interesting only when the state is maintained (e.g.~a \sch-cat state remains a cat state). This is a stronger requirement than just having any nonclassical state, a requirement which cannot be satisfied by our amplifier model.\footnote{Generally, when nonclassicality is important, one may choose more clever ways of amplification. For example, a probabilistic amplifier may be used to boost a cat state if we want to combat photon loss and at the same time and have the state remain as a cat for whatever information processing protocol to be used.} Hence we do not consider nonclassical inputs in this paper. Our objective here is to make clear the breakdown of the small-noise approximation in phase diffusion.
        
    \subsection{Phase diffusion from the Glauber--Sudarshan distribution}
        
        Coherent states are simplest to describe using the $P$ (also called the Glauber--Sudarshan) representation of $\rho(t)$. This is defined by a $P(\alpha,\alpha^*,t)$ such that the state at any time $t$ may be expanded as
        \begin{align}
	        \rho(t) = \int_{\mathbbm{C}} d^2\alpha \; P(\alpha,\alpha^*,t) \op{\alpha}{\alpha}  \;,
        \end{align}
        where the integral runs over the entire complex plane $\mathbbm{C}$. It is well known that $P(\alpha,\alpha^*,t)$ satisfies a \fp\ equation which can be derived from \eqref{DensityOpEOM} \cite{Car02}. It is also well known that the \fp\ equation can be converted to a set of stochastic differential equations for $\alpha(t)$ and $\alpha^*(t)$ whose dynamics sample phase space so as to be consistent with $P(\alpha,\alpha^*,t)$ \cite{Car02,Ris89}. In this paper we adopt the latter approach of stochastic differential equations. That is, instead of considering an evolving $P$ distribution, we consider $\alpha$ and $\alpha^*$ to be changing in time, described by
        \begin{align}
	        d\alpha(t) = {}& \frac{\kminus}{2} \; \alpha(t) \, dt + \rt{\kup} \, dZ(t)  \\
	        d\alpha^*(t) = {}& \frac{\kminus}{2} \; \alpha^*(t) \, dt + \rt{\kup} \, dZ^*(t) \;,
        \end{align}
        where $dZ(t)$ is a complex Wiener increment given by
        \begin{align}
	        \big[dZ(t)\big]^2 = \big[ dZ^*(t) \big]^2 & = 0  \;,  \\
	        dZ^*(t) \, dZ(t) & = dt  \;.
        \end{align}
        However, to describe phase diffusion we ought to change variables from $\alpha$ and $\alpha^*$ to $N$ and $\Phi$, where $\alpha=\rt{N}\exp(i\Phi)$. Our model for phase diffusion is thus given by
        \begin{align}
            \label{dPhi(t)QL}
            d\Phi(t) & = \rt{\frac{\kup}{2N(t)}} \; dV_\Phi(t)  \;, \\
            \label{dN(t)QL}
            dN(t) & = \big[ \kup + \kminus \, N(t) \, \big] \, dt + \sqrt{2\,\kup \, N(t)} \; dV_N(t)  \;.
        \end{align}
        These equations may be derived by converting \eqref{DensityOpEOM} to its equivalent \fp\ equation in terms of the Glauber--Sudarshan distribution \cite{Car02}. For phase diffusion one then needs to transform the \fp\ equation into polar coordinates \cite{BSP89,Lu90}. It is then simple to show that \eqref{dPhi(t)QL} and \eqref{dN(t)QL} are equivalent to the \fp\ equation in polar form. Since we are only interested in amplification, $\kminus$ will be a positive real number. Equations \eqref{dPhi(t)QL} and \eqref{dN(t)QL} are to be interpreted as \ito\ equations with $dV_N(t)$ and $dV_\Phi(t)$ being independent Wiener increments satisfying
        \begin{align}
	        \big[ dV_N(t) \big]^2 = \big[ dV_\Phi(t) \big]^2 & = dt  \;,  \\
	        dV_N(t) \, dV_\Phi(t) & = 0  \;.
        \end{align}
        It is now clear from \eqref{dPhi(t)QL} that the phase $\Phi(t)$ is coupled nonlinearly to $N(t)$. It is essentially this nonlinear coupling to $N(t)$ that makes the phase uncertainty difficult to obtain in closed form. However, as \ito\ equations, \eqref{dPhi(t)QL} and \eqref{dN(t)QL} can be manipulated according to the rules of \ito\ calculus \cite{Gar09}. It is this aspect of our approach that allows us to go further than the small-noise approximation in quantifying the amount of phase noise appearing at the amplifier's output. Existing treatments on the phase uncertainty use \hei\ equations of motion \cite{TGKG92}, the Pegg--Barnett phase operator \cite{BSP89,VP94}, or the phase distribution $\wp_\Phi(\phi,t)$ obtained from quasiprobability distributions \cite{Lu90,Ban91}. While each treatment can be argued to have its own appeal, they all make the small-noise approximation. 

        In order to study the effectiveness of the small-noise approximation we will calculate the time-dependent phase variance
        \begin{align}
        \label{V[Phi]Defn}
	        {\rm V}\big[ \Phi(t) \big] = {\rm E}\big[ \Phi^2(t) \big] - \big\{ {\rm E}\big[ \Phi(t) \big] \big\}^2  \;.
        \end{align}
        The expectation value of any function of phase $f(\Phi)$ at time $t$ is defined formally by
        \begin{align}
        \label{E[f(Phi)]}
	        {\rm E}\big[f\big(\Phi(t)\big)\big] = \int^{\phi_0+2\pi}_{\phi_0} d\phi \; f(\phi) \, \wp_\Phi(\phi,t)  \;,
        \end{align}
        where the phase distribution $\wp_\Phi(\phi,t)$ is defined as the marginal distribution of $\tilde{P}(n,\phi,t)$ (the $P$ function parametrised by $n$ and $\phi$) with $n$ integrated over. Note that we are following the convention of using a capital letter to denote a random variable and the corresponding small letter for its realisation. There are several measures of phase uncertainty that one can choose from \cite{BB93,Hal93}. Thus, given that phase is a cyclic variable one may wonder why we have not used other measures which are better at handling cyclic property of phase. The reason is that the cyclic nature of phase never really shows up here, at least for the simple example of a coherent-state input considered here. We will see that the phase distribution of the amplified field is always a single-peaked function which can always be placed in the centre of an appropriately chosen interval $[\phi_0,\phi_0+2\pi]$. In this case the usual variance as given by \eqref{V[Phi]Defn} can be used \cite{BB93}.

        It is widely known that $P(\alpha,\alpha^*,t)$ can be solved in closed form for Gaussian input states. However, this does not imply that the phase noise can be derived in closed form. This is because on reparametrising $P(\alpha,\alpha^*,t)$ to get $\tilde{P}(n,\phi,t)$, the nonlinear dependence of phase on the photon number gets introduced into the problem. The phase variance then involves calculating the first and second moments with respect to the marginal distribution $\wp_\Phi(\phi,t)$. It is shown in Appendix~D [see \eqref{wpPhi(t)}] that even when restricted to the high-gain limit, the phase distribution $\wp_\Phi(\phi,t)$ is very complicated for a coherent-state input which makes the moments of the phase impossible to derive.
        
\section{The small-noise approximation and beyond}
\label{SmallNoise+InvNumExp}

    The small-noise approximation is first explained in Sec.~\ref{IntroMINexp} and we show how an expression for the phase variance can be obtained under this approximation. This naturally sets the stage for the main method of analysis---the inverse-number expansion in Sec.~\ref{InvNumExp}. The expansion can be stated quite simply and the idea behind it is in fact quite straightforward. Therefore we leave the details of its derivation to Appendix~\ref{Result}.
    
    \subsection{Phase variance within the small-noise approximation}
    \label{IntroMINexp}
    
        To obtain the phase variance we need its second moment. Using the \ito\ chain rule and \eqref{dPhi(t)QL} we obtain
        \begin{align}
        \label{dE[Phi2]/dt}
	        \ddt \; \E\big[ \Phi^2(t) \big] = \frac{\kup}{2} \; \E\bigg[ \frac{1}{N(t)} \bigg]  \;.
        \end{align}
        The second moment of phase thus depends on the expectation value of $\E[1/N(t)]$. As alluded to in Sec.~\ref{Introduction}, the small-noise approximation assumes the field being amplified has a narrow photon-number distribution so that one may regard $N(t)$ as essentially a sure variable (a deterministic quantity) whose value is equal to $\E[N(t)]$. Under this assumption, one may thus replace \eqref{dE[Phi2]/dt} by
        \begin{align}
        \label{LargePhotonApprox}
	        \ddt \; \E\big[ \Phi^2(t) \big] = \frac{\kup}{2\,\E[N(t)]}  \;.
        \end{align}
        We can then derive $\E[N(t)]$ directly from \eqref{dN(t)QL} to be [or take it from \eqref{<N(t)>} with the appropriate change of notation since the photon number is already a normally ordered quantity]
        \begin{align}
        \label{E[N(t)]Exact}
	        \E\big[N(t)\big] = G_t \, \E\big[N(0)\big] + \frac{\kup}{\kminus} \; \big( G_t - 1 \big)  \;,
        \end{align}
        Here we see explicitly that under the small-noise approximation the fluctuations in $N(t)$ are ignored as it is only the mean of $N(t)$ that is used to determine the phase variance. This is clearly quite different to $\E[1/N(t)]$ where both a deterministic part and a noisy part contribute to the denominator inside the expectation value. As we shall see, the inverse-number expansion on the other hand attempts to deal with the $1/N(t)$ dependence wholly which allows the fluctuations in $N(t)$ as given in \eqref{dN(t)QL} to influence the phase variance. We can now derive the phase uncertainty under the small-noise approximation by substituting \eqref{E[N(t)]Exact} into \eqref{LargePhotonApprox}. The derivation is a little tedious so we leave the details to Appendix~\ref{SmallNoiseVar}. The result is simply
        \begin{align}
        \label{Vsmallnoise}
	        {\rm V}[\Phi(t)] = {}& {\rm V}[\Phi(0)] - \frac{1}{2} \, \ln\!\Bigg[ \frac{\E[N(0)]}{\E[N(0)]+\kup/\kminus} \Bigg] \nonumber \\
	                             & - \frac{1}{2} \ln\Bigg[  \frac{\big\{\E[N(0)] + \kup/\kminus \big\} G_t}{\big\{ \E[N(0)] + \kup/\kminus \big\} G_t - \kup/\kminus} \Bigg].
        \end{align}
        Later we will compare the phase variance obtained from the inverse-number expansion to \eqref{Vsmallnoise}.
        
    \subsection{Inverse-number expansion}
    \label{InvNumExp}

\begin{figure}[t]
        \centerline{\includegraphics[width=0.85\textwidth]{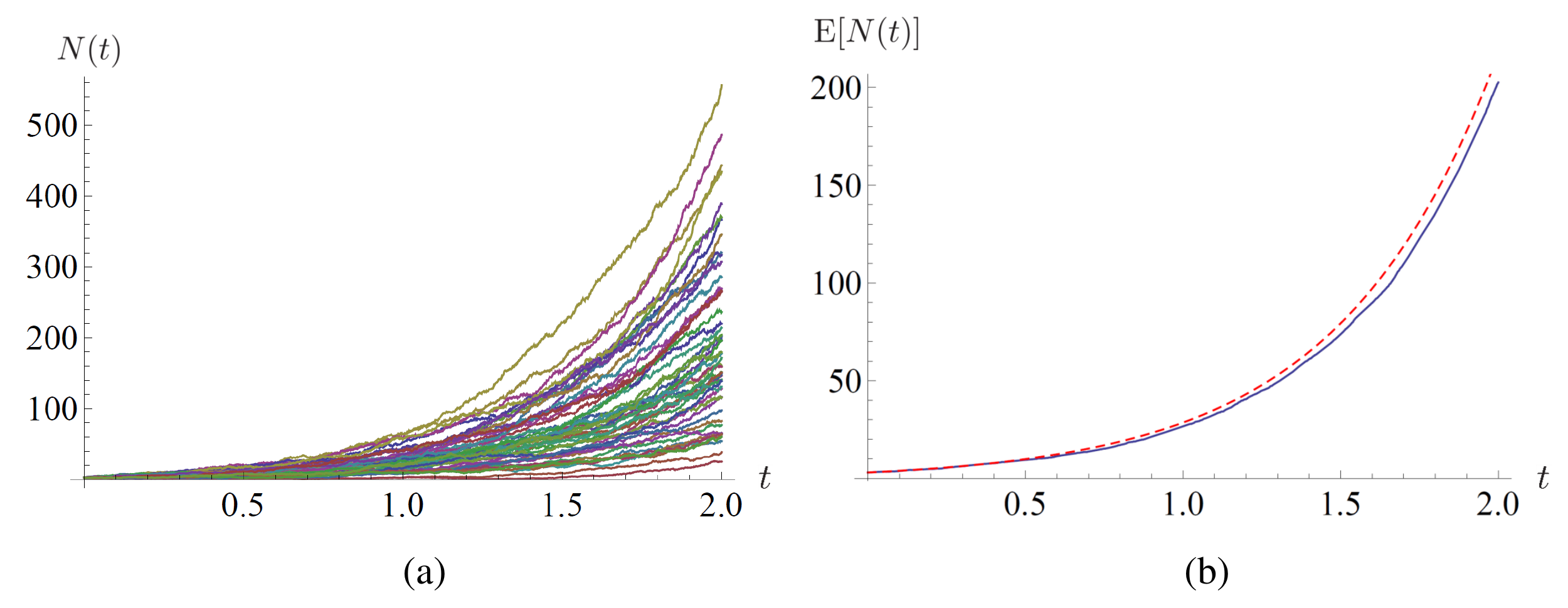}}
        \caption{\label{NumberAmplification} (a) Fifty realisations of $N(t)$ obtained from \eqref{dN(t)QL} with $\kdn=0$ and $\kup=2$ for a coherent-state input $\ket{\alpha}$ with $|\alpha|^2=N(0)=3$. (b) Ensemble average of $N(t)$ obtained from averaging over the realisations in (a) (blue solid curve), and analytically from \eqref{dN(t)QL} (red dashed curve).}
    \end{figure}

    In order to go further than the small-noise approximation we must try to deal with the $\E[1/N(t)]$ appearing in \eqref{dE[Phi2]/dt}. To this end, let us work out how $\E[1/N(t)]$ changes in time. Using \eqref{dN(t)QL} and \ito\ calculus we find
    \begin{align}
    \label{dE[Ups]/dt}
	    \ddt \; \E\bigg[\frac{1}{N(t)}\bigg] = -\,\kminus \, \E\bigg[\frac{1}{N(t)}\bigg] + \kup \; \E\bigg[\frac{1}{N^2(t)}\bigg]   \;.
    \end{align}
    Hence we see that ${\rm E}[1/N(t)]$ depends on ${\rm E}[1/N^2(t)]$. However, since amplification increases the photon number we expect that on average $N(t)$ to increase, and hence $1/N(t)$ to decrease. This is illustrated in Fig.~\ref{NumberAmplification} where different realisations of $N(t)$ corresponding to an ideal linear amplifier are shown. 
    If at time $t$ we find that on average $1/N^2(t) \ll 1/N(t)$, then we can neglect the second term in \eqref{dE[Ups]/dt} and obtain, as a first-order approximation, 
    \begin{align}
    \label{1stIteration}
	    \E\bigg[ \frac{1}{N(t)} \bigg] \approx \frac{1}{G_t} \; \E\bigg[ \frac{1}{N(0)} \bigg]   \;.
    \end{align}
    Clearly, one can continue in a similar fashion and improve on \eqref{1stIteration} by taking into account the effect of $\E[1/N^2(t)]$. The evolution of $\E[1/N^2(t)]$ is simple to obtain and is given by 
    \begin{align}
    \label{dE[Ups2]/dt}
	    \ddt \; \E\bigg[\frac{1}{N^2(t)}\bigg] = & - 2\,\kminus \,\E\bigg[\frac{1}{N^2(t)}\bigg] + 4 \, \kup \, \E\bigg[ \frac{1}{N^3(t)} \bigg]   \;.
    \end{align}
    On neglecting ${\rm E}[1/N^3(t)]$ we can then solve \eqref{dE[Ups2]/dt} straightforwardly. The second-order approximation to $\E[1/N(t)]$ is then obtained by first solving \eqref{dE[Ups2]/dt} with ${\rm E}[1/N^3(t)]$ neglected and then using the resulting expression for ${\rm E}[1/N^2(t)]$ to solve \eqref{dE[Ups]/dt}. Performing these steps we arrive at
    \begin{align}
    \label{E[Ups]2ndOrder}
	    \E\bigg[ \frac{1}{N(t)} \bigg] & \approx \frac{1}{G_t} \; \E\bigg[ \frac{1}{N(0)} \bigg] + \frac{\kup}{\kminus} \;\, \frac{1}{G_t} \: \bigg( \, 1 -  \frac{1}{G_t} \bigg) \, \E\bigg[ \frac{1}{N^2(0)} \bigg]  \;.
    \end{align}
    From this we can now appreciate the essential simplification made by the small-noise approximation---namely that it avoids the coupling of $\E[1/N(t)]$ to higher order moments in $1/N(t)$. In general we will find that $E[1/N^k(t)]$ couples to $E[1/N^{k+1}(t)]$ so that in principle, solving for ${\rm E}[1/N(t)]$ requires us to solve an infinite set of coupled differential equations involving all the moments of $1/N(t)$. In practice however, we would expect (and in fact we shall find) that truncating the set of coupled differential equations at $\E[1/N^{K}(t)]$ by neglecting $\E[1/N^{K+1}(t)]$ for a finite $K$ captures the dynamics of $\E[1/N(t)]$ sufficiently accurately. Our objective is thus to first obtain a general expression for $\E[1/N(t)]$ for any $K$ (i.e.~any truncation) and subsequently an expression for the phase variance ${\rm V}[\Phi(t)]$. It will be convenient to define 
    \begin{align}
	    \Upsilon(t) = \frac{1}{N(t)}  \;,
    \end{align}
    where using \ito's lemma, one can show that $\Upsilon(t)$ is the stochastic process defined by
    \begin{align}
    \label{dUpsMainText}
	    d\Upsilon(t) = & - \big[ \kminus \Upsilon(t) - \kup \, \Upsilon^2(t) \big] \, dt - \Upsilon^2(t) \, \rt{ 2 \, \kup \Upsilon^{-1}(t)} \; dV_N(t)  \;.
    \end{align}
    Based on \eqref{1stIteration} and \eqref{E[Ups]2ndOrder} we propose the following expansion for $\E[\Upsilon(t)]$\,,
    \begin{align}
    \label{E[Ups(t)]KthOrder}
	    \E\big[ \Upsilon(t) \big] = \sum_{n=1}^K \, g_n(t) \, \E\big[ \Upsilon^{n}(0) \big].
    \end{align}
    We call \eqref{E[Ups(t)]KthOrder} the inverse-number expansion up to $K$th order. Equations \eqref{1stIteration} and \eqref{E[Ups]2ndOrder} are thus examples of the inverse-number expansion up to order one, and order two respectively. The first two expansion coefficients can thus be read off from these equations,
    \begin{align}
    \label{g1Explicit}
	    g_1(t) & = \frac{1}{G_t}  \;,  \\
    \label{g2Explicit}
	    g_2(t) & = \frac{\kup}{\kminus} \;\, \frac{1}{G_t} \: \bigg( \, 1 -  \frac{1}{G_t} \bigg)  \;.
    \end{align}
    Note that to avoid convergence issues in \eqref{E[Ups(t)]KthOrder} the realisations of $N(0)$ need to be greater than one. The problem of obtaining the phase variance now requires one to find the expansion coefficients $\{g_n(t)\}_{n=1}^K$ up to any order $K$, and for arbitrary parameter values $\kup$ and $\kdn$. In essence, the important step is the application of \ito's lemma to the $n$th power of $\Upsilon(t)$, giving
    \begin{align}
    \label{dE[Ups(t)]/dt}
	    \frac{d}{dt} \; {\rm E}\big[\Upsilon^n(t)\big] = b_n \, {\rm E}\big[\Upsilon^n(t)\big] + c_n \; {\rm E}\big[\Upsilon^{n+1}(t)\big]  \;,
    \end{align}
    where we have defined for ease of writing
    \begin{align}
    \label{Coeffbc}
	    b_n = - n\,\kminus \;,  \quad   c_n = n^2\,\kup   \;.
    \end{align}
    We refer the reader to Appendix~\ref{Result} for the proof of \eqref{dE[Ups(t)]/dt} and the subsequent derivation of the time-dependent expansion coefficients. The result is
    \begin{align}
    \label{gn(t)MainText}
	    g_n(t) = \sum_{k=1}^n \, \beta_{k,n} \, e^{b_k\;\!t}, 
    \end{align}
    where $\beta_{1,1} = 1$, otherwise
    \begin{align}
    \label{BetaMainText}
	   \beta_{k,n} = \left. c_1 \, c_2 \, \cdots \, c_{n-1} \middle/ \underset{j\ne k}{\prod_{j=1}^{n}}\,(b_k-b_j) \quad n \ge 2 \right.  \;.
    \end{align}
    Note from \eqref{gn(t)MainText} that $g_n(t)\longrightarrow0$ as $t\longrightarrow\infty$. This means that the right-hand side of \eqref{dE[Phi2]/dt} eventually goes to zero and the second moment of phase reaches a constant value. We can write out, as an example, the first three coefficients of the inverse-number expansion:
    \begin{widetext}
    \begin{align}
    \label{g1MainText}
	    g_1(t) = {}& e^{b_1\;\!t} \;, \\  
    \label{g2MainText}
	    g_2(t) = {}& \frac{c_1}{(b_1-b_2)} \; e^{b_1\;\!t} + \frac{c_1}{(b_2-b_1)} \; e^{b_2\;\!t}  \;, \\
    \label{g3MainText}
	    g_3(t) = {}& \frac{c_1 \, c_2}{(b_1-b_2)(b_1-b_3)} \, e^{b_1\;\!t} + \frac{c_1 \, c_2}{(b_2-b_1)(b_2-b_3)} \; e^{b_2\;\!t} 
	             + \frac{c_1\,c_2}{(b_3-b_1)(b_3-b_2)} \; e^{b_3\;\!t} \;.
    \end{align}
    \end{widetext}
    It is simple to see that \eqref{g1Explicit} and \eqref{g2Explicit} are reproduced on substituting \eqref{Coeffbc} into \eqref{g1MainText} and \eqref{g2MainText}. The advantage of the inverse-number expansion can also be seen in the complexity of $g_3(t)$, where it would have been somewhat tedious to derive by solving a set of three coupled differential equations. Of course, after obtaining \eqref{E[Ups(t)]KthOrder} one still has to calculate $\E[\Phi^2(t)]$ according to \eqref{dE[Phi2]/dt}, and then the variance defined by \eqref{V[Phi]Defn}. The expression for $\E[\Phi^2(t)]$ thus entails an integral of $g_n(t)$. We provide a discussion on the validity of the inverse-number expansion in Appendix~\ref{Validity}.
    
\section{Phase uncertainty}
\label{PhaseUncert}

    We now have everything to present our main result. It is shown that for relatively weak inputs (whose number average is a few photons), the small-noise approximation inadequately captures the high-gain output phase noise even for an ideal linear amplifier. Such a linear amplifier is optimal with respect to the small-noise approximation, so if the approximation fails here, one can expect it to fail more severely for either nonideal linear amplifiers, or ideal amplifiers but with less gain. We have also provided a discussion on the use of the Glauber--Sudarshan distribution in describing the phase of the amplified field in Appendix~\ref{PhaseWithinP}. In essence this validates our result in the high-gain regime.

    \subsection{General expression}
    
    Substituting \eqref{E[Ups(t)]KthOrder} back into \eqref{dE[Phi2]/dt} and integrating we have
    \begin{align}
    	{\rm E}\big[\Phi^2(t)\big] = {\rm E}\big[\Phi^2(0)\big] + \frac{\kup}{2} \, \sum_{n=1}^K  \, \E\big[ \Upsilon^{n}(0) \big]  \int^t_0 dt' \, g_n(t')    \quad.
    \end{align}
    To obtain the variance of $\Phi(t)$ we also require its mean. This is simple to get from \eqref{dPhi(t)QL} as the phase is entirely driven by a Wiener process so that on averaging \eqref{dPhi(t)QL} we have
    \begin{align}
    \label{E[Phi]}
	    {\rm E}\big[ \Phi(t) \big] = {\rm E}\big[ \Phi(0) \big]   \;.
    \end{align}
    Squaring \eqref{E[Phi]} and recalling the definition of the variance from \eqref{V[Phi]Defn} we arrive at
    \begin{align}
    \label{V[Phi(t)]KthOrder}
	    {\rm V}\big[ \Phi(t) \big] = {\rm V}\big[\Phi(0)\big] +  \sum_{n=1}^K  \,  \chi_n(t) \, \E\big[ \Upsilon^{n}(0) \big]   \;.
    \end{align}
    where we have defined, on using \eqref{gn(t)MainText},
    \begin{align}
	    \chi_n(t) = \frac{\kup}{2} \, \int^t_0 dt' \, g_n(t') = \frac{\kup}{2} \, \sum_{k=1}^n \; \frac{\beta_{k,n}}{b_k} \; \big( e^{b_k\;\!t} - 1 \big)  \;.
    \end{align}
    
    \subsection{Ideal linear amplifier}

    As explained earlier, an ideal linear amplifier is an amplifier that adds the least amount of noise. Such an amplifier corresponds to the model in Sec.~\ref{Model} with $\kdn=0$ so that 
    \begin{align}
	    \kminus = \kup  \;.
    \end{align}
    Thus the ideal linear amplifier is entirely characterised by a single parameter, $\kup$ (assuming a fixed $t$). This sufficiently simplifies the expression for the phase variance so that we may write it as an explicit function of $\kup$. From \eqref{Coeffbc} we have
    \begin{align}
	    b_n & = - n\,\kup,  \\	
	    c_1 \, c_2 \, \cdots \, c_{n-1} & = \big[(n-1)!\big]^2 \; (\kup)^{n-1}  \;,  \\ 
	    \underset{j\ne k}{\prod_{j=1}^{n}}\,(b_k-b_j) & = (\kup)^{n-1} \, \underset{j\ne k}{\prod_{j=1}^{n}}\,(j-k) \;. 
    \end{align}
    It then follows that $\beta_{k,n}$ is independent of $\kup$. Recall that we defined $\beta_{1,1}=1$ while
    \begin{align}
	    \beta_{k,n} = \left. \big[\,(n-1)!\,\big]^2 \middle/ \underset{j\ne k}{\prod_{j=1}^{n}}\,(j-k) \;, \quad n \ge 2 \right.  \;,
    \end{align}
    and the phase variance then simplifies to
    \begin{align}
        \label{IdealLinearAmpV[Phi(t)]}
	    {\rm V}\big[ \Phi(t) \big] = {\rm V}\big[\Phi(0)\big] + \sum_{n=1}^K  \,  \chi_n(t) \, \E\big[ \Upsilon^{n}(0) \big]   \;,
    \end{align}
    where $\chi_n(t)$ is now given by
    \begin{align}
    \label{Xn(t)}
	    \chi_n(t) = \frac{\big[\,(n-1)!\,\big]^2}{2} \, \sum_{k=1}^n \,  \frac{1-G_t^{-k}}{k\,\underset{j\ne k}{\prod_{j=1}^{n}}\,(j-k)}.
    \end{align}
    Note that the product in the denominator of \eqref{Xn(t)} skips over terms for which $j=k$ so that for $k=1$ we have $\prod_{j=1,j\ne1}^{1}\,(j-1)=1$. The first time-dependent coefficient for the ideal linear amplifier is therefore
    \begin{align}
	    \chi_1(t) = \frac{1}{2} \; \bigg( 1 - \frac{1}{G_t} \bigg)  \;.
    \end{align}
    
    \begin{figure}[t]
        \centerline{\includegraphics[width=0.8\textwidth]{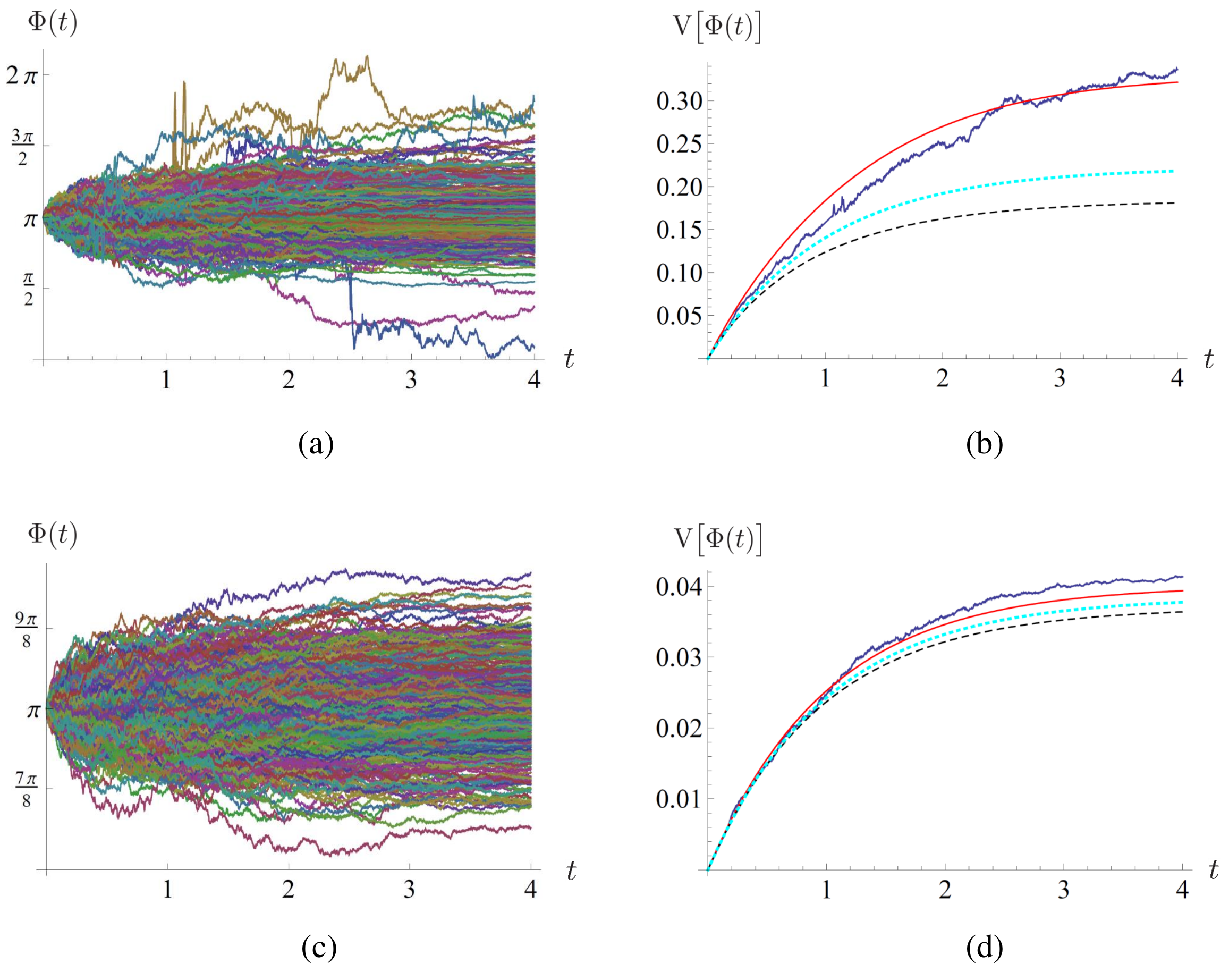}}
        \caption{\label{SmallNoiseComparisonIdealLinearAmplifier} Phase variance for an ideal linear amplifier with $\kup=1$ and $\Phi(0)=\pi$. Comparison between the `exact' phase variance derived using the inverse-number expansion for $K=4$ versus the small-noise approximation and a first-order inverse-number expansion. The analytic expression for phase variance is shown to be consistent with the sample variance corresponding to 500 realisations of $\Phi(t)$. (a) and (b): $N(0)$=2.25. (c) and (d): $N(0)$=13.}
    \end{figure}
    Let us now compare the phase variance obtained from integrating the inverse-number expansion to the variance obtained with the small-noise approximation [recall \eqref{Vsmallnoise} from Sec.~\ref{IntroMINexp}]. This is illustrated in Fig.~\ref{SmallNoiseComparisonIdealLinearAmplifier} for an ideal linear amplifier which we now explain. First, in  Fig.~\ref{SmallNoiseComparisonIdealLinearAmplifier}(a) we show 500 sample paths of the phase which gives us a visualisation of how the phase of an input signal diffuses as it gets amplified. In Fig.~\ref{SmallNoiseComparisonIdealLinearAmplifier}(b) we plot the phase variance from the 500 phase realisations (blue jiggly curve) along with the variance obtained from \eqref{IdealLinearAmpV[Phi(t)]} for $K=4$ (red, solid line), $K=1$ (cyan, dotted line), and the small-noise approximation (black dashed line). Recall that we defined the small-noise approximation in \eqref{LargePhotonApprox} back in Sec.~\ref{IntroMINexp}. The same is true for Fig.~\ref{SmallNoiseComparisonIdealLinearAmplifier}(c) and (d) with the only difference being the value of $N(0)$. For Fig.~\ref{SmallNoiseComparisonIdealLinearAmplifier}(a) and (b) we took $N(0)=2.25$, while for (c) and (d) we used $N(0)=13$. This allows us to see how the small-noise approximation performs when applied to the amplification of two inputs of different strengths. Clearly, the inverse-number expansion does a much better job of estimating the output phase noise for relatively weak inputs. Another interesting observation from Fig.~\ref{SmallNoiseComparisonIdealLinearAmplifier} is that even the lowest-order approximation ($K=1$ in the inverse-number expansion) as explained in Sec.~\ref{InvNumExp} performs visibly better than the small-noise approximation.
    
    With the comparison shown in Fig.~\ref{SmallNoiseComparisonIdealLinearAmplifier} at hand, let us now try to understand some of its essential features: First note how in both Fig.~\ref{SmallNoiseComparisonIdealLinearAmplifier}(b) and (d) the small-noise approximation appears to underestimate ${\rm V}[\Phi(t)]$. It is in fact a general property that the small-noise approximated phase variance provides a lower bound for ${\rm V}[\Phi(t)]$. We can show this by considering the evolution of the variance given by 
    \begin{align}
        \label{dV[Phi]/dt}
	    \ddt \; {\rm V}[\Phi(t)] = \ddt \; \E[\Phi^2(t)] = \frac{\kup}{2} \; \E\bigg[\frac{1}{N(t)}\bigg]  \;,
    \end{align}
    where we have used the fact that $\E[\Phi(t)]$ is constant. Now we note that Jensen's inequality states that $\E[f(X)]\ge f(\E[X])$ where $f(X)$ is a convex function of the random variable $X$. It can then be shown that $1/N$ is convex and hence it follows from Jensen's inequality that
    \begin{align}
    \label{dV/dtIneq}
	    \ddt \; {\rm V}[\Phi(t)] \ge \frac{\kup}{2} \; \frac{1}{\E[N(t)]}  \;.
    \end{align}
    The right-hand side of the inequality \eqref{dV/dtIneq} is exactly the rate of change of the phase variance in the small-noise approximation. Thus the exact phase variance must always rise at least as fast, and hence will be greater than or equal to, the phase variance in the small-noise approximation. Note that \eqref{dV[Phi]/dt} also says that as $N(t)$ increases without bound, 
    \begin{align}
	    \lim_{t\to\infty} \; \ddt \; {\rm V}[\Phi(t)] = 0  \;.
    \end{align}
    This means that the phase variance approaches to some constant value and is therefore upper bounded. This precisely is what we see in both Fig.~\ref{SmallNoiseComparisonIdealLinearAmplifier}(b) and (d) where the phase variance can be seen to level off after some transient dynamics. 
    
\begin{figure}[t]
        \centerline{\includegraphics[width=0.45\textwidth]{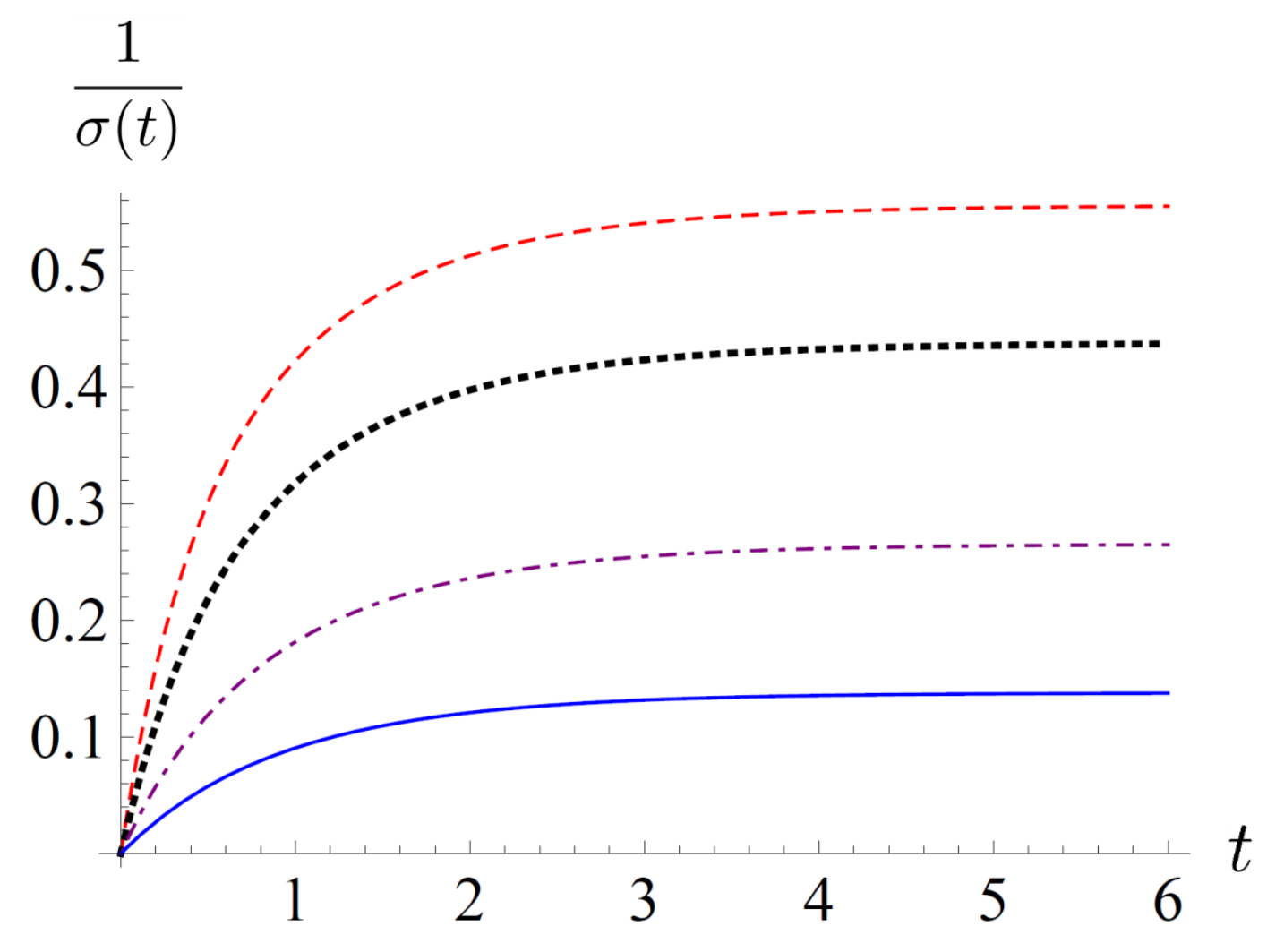}}
        \caption{\label{InverseSNR} Inverse of the signal-to-noise ratio of an ideal amplifier for different the input strengths: $N(0)=2$ (red dashed curve), $N(0)=3$ (black dotted curve), $N(0)=6$ (purple, dash-dot curve), and $N(0)=13$ (blue solid curve). }
    \end{figure}

    We can also try to understand why the small-noise approximation performs worse for smaller inputs. Recall from Sec.~\ref{IntroMINexp} that we expect the small-noise approximation to be good when the photon-number distribution is narrow. If the phase variance obtained from making such an approximation is to be accurate for all times then this condition must be met for all times as well. The condition of a narrow number distribution can also be understood as having a strong signal-to-noise ratio $\sigma(t)$, defined as 
    \begin{align}
	    \sigma(t) = \frac{\big(\,\E[N(t)]\,\big)^2}{{\rm V}[N(t)]}  \;.
    \end{align}
    In fact, a Taylor expansion around the mean of $N(t)$ up to second order gives
    \begin{align}
    \label{ApproxE[1/N]}
	    \E\bigg[ \frac{1}{N(t)} \bigg] \approx \frac{1}{\E[N(t)]} \; \Bigg[ 1 + \frac{1}{\sigma(t)} \Bigg]   \;.
    \end{align}
    Therefore to understand why the discrepancy between the ${\rm V}[\Phi(t)]$ obtained from the small-noise approximation and inverse-number expansion is larger in Fig.~\ref{SmallNoiseComparisonIdealLinearAmplifier}(b) compared to Fig.~\ref{SmallNoiseComparisonIdealLinearAmplifier}(d), we should consider how $1/\sigma(t)$ evolves for different $N(0)$. From \eqref{ApproxE[1/N]} we expect the larger $1/\sigma(t)$ is, the worse the small-noise approximation performs. A closed-form expression for $1/\sigma(t)$ can be derived. Remember from \eqref{E[N(t)]Exact} of Sec.~\ref{IntroMINexp} we already have (for any $\kup$ and $\kdn$) 
    \begin{align}
    \label{E[N(t)]}
	    \E\big[N(t)\big] = G_t \,\E\big[N(0)\big] + \frac{\kup}{\kminus} \; \big( G_t - 1 \big)  \;.
    \end{align}
    So to obtain $\sigma(t)$ we only need $\E[N^2(t)]$. It is not too difficult to show that in general (i.e.~arbitrary $\kup$ and $\kdn$),
    \begin{align}
    \label{E[N2(t)]}
	    \E[N^2(t)] = G^2_t \, \E[N^2(0)] 
	                 + 4 \kup \, G^2_t \, \bigg\{ \bigg[ \frac{\kminus \, \E[N(0)]+\kup}{\kminus^2} \bigg] \bigg( 1 - \frac{1}{G_t} \bigg)
	                 - \frac{\kup}{2\kminus^2} \bigg( 1 - \frac{1}{G^2_t} \bigg) \bigg\}.
    \end{align}
    The number variance ${\rm V}[N(t)]= \E[N^2(t)]-(\E[N(t)])^2$ is thus given explicitly by \eqref{E[N(t)]} and \eqref{E[N2(t)]}, and hence $\sigma(t)$. The analytic result for $1/\sigma(t)$ is plotted in Fig.~\ref{InverseSNR} to illustrate its dependence on $N(0)$. In particular, the $N(0)=2$ and $N(0)=13$ curves correspond to the two situations shown in Fig.~\ref{SmallNoiseComparisonIdealLinearAmplifier} (ignoring the small difference of 0.25 for the weaker input). We have also plotted $1/\sigma(t)$ for intermediate values of $N(0)$ to illustrate how the evolution of the signal-to-noise ratio depends on $N(0)$.
    
    As can be seen from Fig.~\ref{InverseSNR}, $1/\sigma(t)$ rises much faster for $N(0)=2$ than $N(0)=13$. This is consistent with the phase variance observed in Fig.~\ref{SmallNoiseComparisonIdealLinearAmplifier}(b) where the small-noise phase variance diverges from the inverse-number expansion result early on during transience and severely  underestimates the stationary phase variance thereafter. On the other hand, the $1/\sigma(t)$ curve for $N(0)=13$ rises much more slowly and remains at a much lower level compared to the $N(0)=2$ case. Thus we now understand why in Fig.~\ref{SmallNoiseComparisonIdealLinearAmplifier}(d) the small-noise phase variance stays close to the inverse-number formula during transience and continues to do so at later times when compared to Fig.~\ref{SmallNoiseComparisonIdealLinearAmplifier}(b). 
    
\begin{figure}[t]
        \centerline{\includegraphics[width=0.8\textwidth]{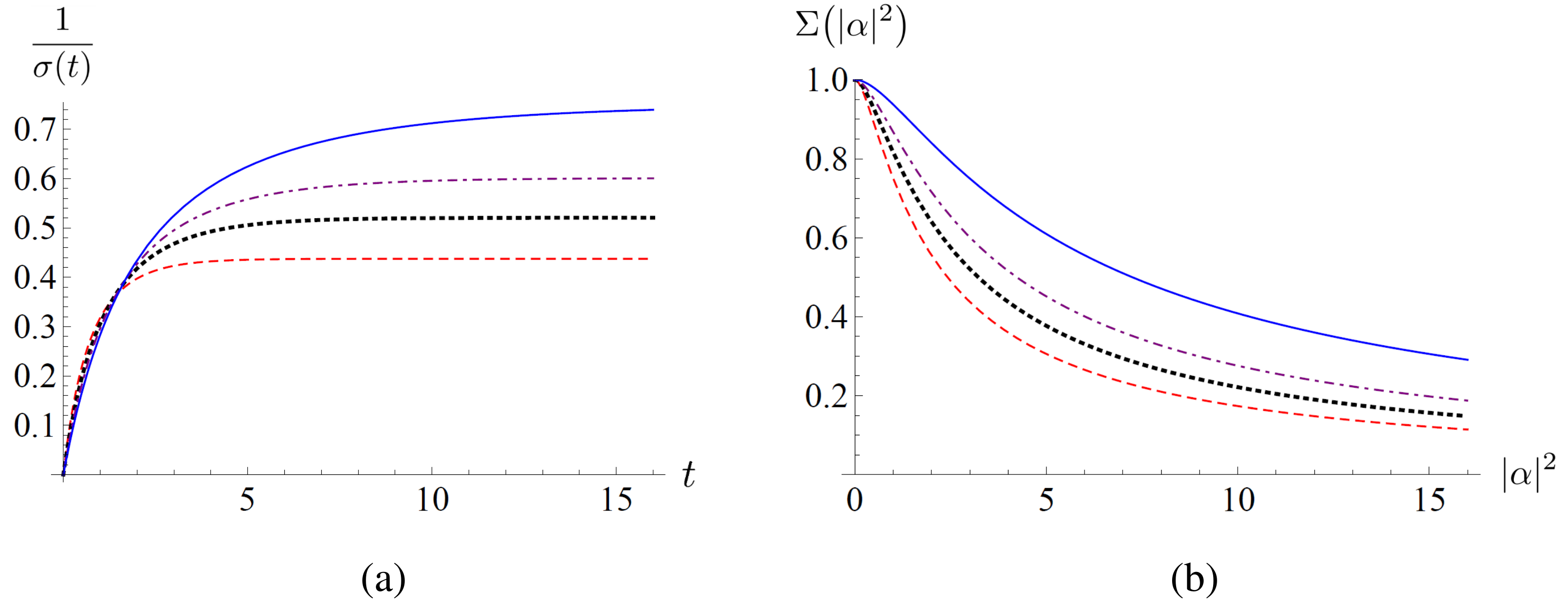}}
        \caption{\label{HighGainInverseSNR}(a) Inverse of the signal-to-noise ratio for $N(0)=3$ but different levels of non-ideality: $\kup=0.6$, $\kdn=0.4$ (blue solid curve), $\kup=0.7$, $\kdn=0.3$ (purple, dash-dot curve), $\kup=0.8$, $\kdn=0.2$ (black dotted curve), $\kup=1$, $\kdn=0$ (red dashed curve). Note that for short times ($t\lesssim1.5$) i.e.~low $G_t$, the order of performance is actually reversed (not clearly visible from the plot). (b) Inverse of the signal-to-noise ratio for large gain as a function of the input strength. The colour coding corresponding to different $\kup$ and $\kdn$ values is the same as in (a).}
    \end{figure}

    It is also interesting to consider the case of a fixed input strength for different levels of non-ideality. This is shown in Fig.~\ref{HighGainInverseSNR}(a) for $N(0)=3$. For nonideal amplifiers, $1/\sigma(t)$ takes longer to reach steady state value so we have plotted the curves for a longer time. At long times (i.e.~large $G_t$), the more ideal the amplifier is, the lower its $1/\sigma(t)$ value. This ordering is in fact preserved for all input strengths for large $G_t$. With the help of \eqref{E[N(t)]} and \eqref{E[N2(t)]}, we obtain a simple expression for $1/\sigma(t)$ when $G_t\gg1$ (assuming a coherent-state input with amplitude $\alpha$)
    \begin{align}
    \label{1/sigmaFormula}
	    \frac{1}{\sigma} = \Sigma\big(|\alpha|^2\big) = \frac{\big(|\alpha|^2+\kup/\kminus\big)^2 - |\alpha|^4}{\big(|\alpha|^2+\kup/\kminus\big)^2}.
    \end{align}
    From this we see that $\Sigma(|\alpha|^2)\le 1$ (with equality when $|\alpha|^2=0$ i.e.~no input), and that $\Sigma(|\alpha|^2)\longrightarrow0$ for $|\alpha|^2\longrightarrow\infty$, so that nonideal and ideal amplifiers become indistinguishable in so far as $\sigma(t)$ is concerned. For all nonzero and finite $|\alpha|^2$ the behaviour of $\Sigma(|\alpha|^2)$ is shown in Fig.~\ref{HighGainInverseSNR}(b). We plot $\Sigma(|\alpha|^2)$ for the same nonideal amplifiers as in Fig.~\ref{HighGainInverseSNR}(a). We find the more ideal the amplifier is, the better it performs for any input strength. This again corroborates with discrepancy seen in Fig.~\ref{SmallNoiseComparisonIdealLinearAmplifier} between the phase variance obtained from the small-noise approximation and the inverse-number expansion.
    
\section{Conclusion}
\label{Conclusion}

In this paper we have shown the often-used small-noise approximation to be inadequate for capturing phase diffusion in linear amplifiers when the input contains a few photons on average. However, the phase noise of a `slightly' stronger input (say ten photons or more) may be described reasonably well within the small-noise approximation. We have demonstrated this even under conditions most favourable to the small-noise approximation---amplification of a single-mode field by an ideal linear amplifier with large gain (equivalent to amplifying an input for a long time with a fixed amplification rate). This is also the regime where one would ideally like to operate a linear amplifier. We obtained such a result by proposing the inverse-number expansion using the $P$ (or Glauber--Sudarshan) function. From our examples, we saw that the inverse-number expansion with a reasonably small set of terms was sufficient to capture the phase uncertainty beyond transience. In terms of the $P$ function, a coherent state is a completely noise-free point in phase space which makes such inputs extremely easy to handle. In this case, all of noise seen at the amplifier's output is interpreted as added noise due to the amplifier being quantum-mechanical. The disadvantage, is that it cannot predict the correct transient phase noise (though this is actually the less interesting case). The extended discussion of Appendix~\ref{PhaseWithinP} implies that the $W$ (i.e.~Wigner) function is a better choice for capturing phase diffusion because it has the special property of giving true marginals. One possible extension which can be explored is to recast the inverse-number expansion in terms of the $W$ function. This would then allow us to obtain a valid estimate for the phase noise even at small gain, or short times. The cost to this extension is that the initial inverse-number statistics become much more difficult to calculate even for a simple input such as the coherent state. 

\section*{Acknowledgements}

This research is supported by the MOE grant number RG 127/14, and the National Research Foundation, Prime Minister's Office, Singapore under its Competitive Research Programme (CRP Award No. NRF-CRP-14-2014-02). We would also like to thank Ranjith Nair for useful discussions.

\onecolumn\newpage
\appendix

\section{Phase variance within the small-noise approximation}
\label{SmallNoiseVar}

    We have already said in Sec.~\ref{IntroMINexp} that the small-noise approximation replaces the field intensity by its mean value so that we have for the second moment of $\Phi(t)$,
    \begin{align}
    \label{LargePhotonApproxAppendix}
	    \ddt \; \E\big[ \Phi^2(t) \big] = \frac{\kup}{2\,\E[N(t)]}  \;,
    \end{align}
    where $\E[N(t)]$ is given by 
    \begin{align}
    \label{E[N(t)]ExactAppendix}
	    \E\big[N(t)\big] = \E\big[N(0)\big] \; e^{\kminus\,t} + \frac{\kup}{\kminus} \; \big( e^{\kminus\,t} - 1 \big)  \;.
    \end{align}
    This is a good approximation provided that $(\E[N(t)])^2 \gg V[N(t)]$. Here we shall derive the phase variance in closed form within the small-noise approximation.

    To start, we note that the phase is coupled to $N(t)$ but it is purely noisy (i.e.~it has no deterministic evolution) driven by a Wiener process. The average of $\Phi(t)$ is therefore equal to its initial value $\E[\Phi(0)]$. Integrating \eqref{LargePhotonApproxAppendix} then gives us
    \begin{align}
        \label{Vformal}
	    {\rm V}[\Phi(t)] = \E[\Phi^2(t)] - \big(\E[\Phi(t)]\big)^2 = {\rm V}[\Phi(0)] + \frac{\kup}{2} \int^t_0 dt' \; \frac{1}{\E[N(t')]}  \;.
    \end{align}
    Using \eqref{E[N(t)]ExactAppendix}, we write the integral of $1/\E[N(t)]$ in a more compact form 
    \begin{align}
	    \int^t_0 dt' \; \frac{1}{\E[N(t')]} = \int^t_0 dt' \; \frac{1}{a\,e^{\kminus t'} + b}  \;,
    \end{align}
    where $a$ and $b$ are constants defined by
    \begin{align}
    \label{aConst}
    	a = {}& \E[N(0)] + \frac{\kup}{\kminus} \;,  \\
    \label{bConst}
	    b = {}& -\frac{\kup}{\kminus} \;.
    \end{align}
    The exponential in the denominator can be dealt with if we make the substitution
    \begin{align}
	    u = a \, e^{\kminus \,t}  \;.
    \end{align}
    The corresponding indefinite integral can then be written entirely in terms of $u$:
    \begin{align}
	    \int dt \; \frac{1}{a\,e^{\lambda t} + b} = {}& \frac{1}{\kminus} \int du \; \frac{1}{u(u+b)}  \\
                                            = {}& \frac{1}{\kminus \,b} \; \bigg[ \int du \; \frac{1}{u} - \int du \; \frac{1}{(u+b)} \bigg]  \\
                                            = {}& \frac{1}{\kminus \,b} \; \Big[ \ln u - \ln(u+b) \Big]  \;.
    \end{align}
    The definite integral then follows on plugging in the definition of $u$ and then evaluating the limits. The result is simply
    \begin{align}
	    \int^t_0 dt \; \frac{1}{a\,e^{\lambda t} + b} = {}& \frac{1}{\kminus\,b} \; \Big[ \ln\!\big(a\,e^{\kminus\,t}\big) - \ln\!\big(a\,e^{\kminus\,t}+b\big) \Big]  
	                                                    - \frac{1}{\kminus\,b} \; \Big[ \ln\!\big(a\big) - \ln\!\big(a+b\big) \Big]  \;.
    \end{align}
    Substituting this back into \eqref{Vformal} and using \eqref{aConst} and \eqref{bConst} for $a$ and $b$ we arrive at \eqref{Vsmallnoise} in the main text.
    
\section{Derivation of the inverse-number expansion}
\label{Result}

    \subsection{Main proof}

    Recall from \eqref{dUpsMainText} of Sec.~\ref{InvNumExp} that the stochastic differential equation for $\Upsilon(t)$ is 
    \begin{align}
    \label{dUpsApp}
	    d\Upsilon(t) = - \big[ \kminus \Upsilon(t) - \kup \, \Upsilon^2(t) \big] \, dt - \Upsilon^2(t) \, \rt{ 2 \, \kup \Upsilon^{-1}(t)} \; dV_N(t)  \;.
    \end{align}
    Applying \ito's lemma to $\Upsilon^n(t)$ we have
    \begin{align}
	    d\Upsilon^n(t) = {}& \frac{d\Upsilon^n}{d\Upsilon} \; d\Upsilon(t) + \frac{1}{2} \, \frac{d^2\Upsilon^n}{d\Upsilon^2} \; \big[d\Upsilon(t)\big]^2  \\
                 = {}& - n \Upsilon^{n-1}(t) \, \Big\{ \big[ \kminus \Upsilon(t) - \kup \,\Upsilon^2(t) \big]\,dt 
                       + \Upsilon^2 \,\rt{ 2\,\kup \Upsilon^{-1}(t)} \; dV_N(t) \Big\} \nonumber \\ 
                    & + n(n-1) \, \kup \, \Upsilon^{n+1}(t) \,  dt  \\
    \label{dUpsPowern}                     
	               = {}& -n \, \kminus \, \Upsilon^n(t) \, dt + n^2 \, \kup \Upsilon^{n+1}(t) \, dt - \Upsilon^2 \,\rt{ 2\,\kup \Upsilon^{-1}(t)}  dV_N(t)   \;.  
    \end{align}
    It then follows on taking the average of \eqref{dUpsPowern},
    \begin{align}
    \label{dE[Upsn]/dt}
	    \frac{d}{dt} \; {\rm E}\big[\Upsilon^n(t)\big] = b_n \, {\rm E}\big[\Upsilon^n(t)\big] + c_n \; {\rm E}\big[\Upsilon^{n+1}(t)\big]  \;.
    \end{align}
    where
    \begin{align}
    \label{Appbncn}
	    b_n = - n\,\kminus \;,  \quad    c_n = n^2\,\kup  \;.
    \end{align}
    This proves that ${\rm E}[\Upsilon^n]$ couples only to ${\rm E}[\Upsilon^{n+1}]$. Let us now define a $K\times1$ vector $\bm{x}(t)$ by
    \begin{align}
	    \bm{x}(t) = \fbo{x_1}{x_2}{\vdots}{x_K} = \fbo{{\rm E}[\Upsilon(t)]}{{\rm E}[\Upsilon^2(t)]}{\vdots}{{\rm E}[\Upsilon^K(t)]}  \;.
    \end{align}
    This defines the mean of the inverse-number $\Upsilon$ as the first component of $\bm{x}(t)$. We can then write \eqref{dE[Upsn]/dt} in the form of a linear matrix differential equation for $\bm{x}$:
    \begin{align}
    \label{dx/dt}
	       \ddt \, \bm{x}(t) = A \, \bm{x}(t)  \;,
    \end{align}
    where $A$ is the $K\times K$ matrix
    \begin{align}	
	    A = \left[ \begin{array}{ccccc}   b_1    &  c_1 &     0    &  \cdots   &     0     \\
	                                   0     &  b_2 &    c_2   &           &           \\
	                                 \vdots  &      &  \ddots  &  \ddots   &           \\
	                                         &      &          &  b_{K-1}  &  c_{K-1}  \\
	                                   0     &      &          &           &    b_K    \end{array} \right],
    \end{align}
    Furthermore, $A$ is a triangular matrix with $K$ distinct eigenvalues given directly by $\{b_n\}_{n=1}^K$. Letting $D={\rm diag}(\bm{b})$, an eigendecomposition of $A$ then gives
    \begin{align}
	    A = S D S^{-1} \;, 
    \end{align}
    where the $n$th column of $S$ is the eigenvector corresponding to $b_n$:
    \begin{align}
	    S = \Big[ \bm{v}^{(1)} \; \bm{v}^{(2)} \; \cdots \; \bm{v}^{(K)} \Big]  \;, \quad  	A \, \bm{v}^{(n)} = b_n \, \bm{v}^{(n)}  \;,  \quad  n=1,2,\ldots,K  \;.
    \end{align}
    The solution to \eqref{dx/dt} can then be written in component form as
    \begin{align}
	    x_r(t) = {}& \sum_{j=1}^K \sum_{m=1}^K \sum_{n=1}^K  \, S_{r,j} \, \big(e^{Dt}\big)_{j,m} \, \big(S^{-1}\big)_{m,n} \, x_n(0)  \\
         = {}& \sum_{m=1}^K \sum_{n=1}^K  \, S_{r,j} \, \big(S^{-1}\big)_{m,n} \, e^{b_m\,t} \, \bm{x}_n(0)  \\
    \label{xk(t)}       
         = {}& \sum_{n=1}^K  \Bigg[ \sum_{m=1}^K  \, S_{r,m} \, \big(S^{-1}\big)_{m,n} \, e^{b_m\,t} \Bigg] \, x_n(0)  \;.
    \end{align}
    The inverse-number expansion is then given by \eqref{xk(t)} for $r=1$. This gives us a general expression for the time-dependent coefficients in terms of $S$ and its inverse on comparison to \eqref{E[Ups(t)]KthOrder}:
    \begin{align}
    \label{gn}
	    g_n(t) = \sum_{k=1}^K  \, \beta_{k,n} \, e^{b_k\,t} \;,   \quad  \beta_{k,n} = S_{1,k} \, \big(S^{-1}\big)_{k,n}  \;.
    \end{align}
    It can be shown that $S$ is given by
    \begin{align}
    \label{Smn}
	    S_{m,n} = \begin{cases} 
	           \text{\Large$\frac{\prod_{j=m}^{n-1}c_j}{\prod_{j=m}^{n-1} (b_n-b_j)}$} \;,  \quad m < n  \;,  \\
	           1 \;,  \quad m = n  \;,  \\
	           0 \;,  \quad m > n  \;.
	           \end{cases}
    \end{align}
    Note that $S$ is not an orthogonal matrix so that its transpose does not equal its inverse. Nevertheless, because $S$ is an upper-triangular matrix it is fairly straightforward to analytically obtain its inverse using standard symbolic software such as Mathematica. The inverse of $S$ is again upper triangular, given by
    \begin{align}
    \label{InvSmn}
	    \big(S^{-1}\big)_{m,n} = \begin{cases} 
	                         \text{\Large$\frac{\prod_{j=m}^{n-1}c_j}{\prod_{j=m+1}^{n} (b_m-b_j)}$} \;,  \quad m < n  \;,  \\
	                         1 \;,  \quad m = n  \;,  \\
	                         0 \;,  \quad m > n  \;.
	                         \end{cases}
    \end{align}
    We prove in Appendix~\ref{DerivationOfS} that $S$ is given by \eqref{Smn}. It is interesting to note that $S$ has a structure such that for any $K$, its inverse can be constructed directly by taking matrix powers. This is due to $S$ being an upper-triangular matrix with its main diagonal given by only ones. We can then write it in terms of a nilpotent matrix $N$ as $S=I+N$ where $I$ is the $K \times K$ identity matrix and $N$ is strictly upper triangular such that $N^p=0$ for all $p\ge K$. The inverse of $S$ can then be obtained as a finite (geometric) series in $N$
    \begin{align}
	    S^{-1} = \sum_{q=0}^{K-1} \, N^q  \;.
    \end{align}
    This provides an alternative way to compute $S^{-1}$. Since $N$ is strictly upper triangular, its powers are not too difficult to compute for small values of $K$. Note that even though we need only the first row of $S$ for $g_n(t)$, the remaining elements of $S$ are still required to obtain $S^{-1}$. Taking the first row of $S$ we have 
    \begin{align}	
        S_{1,1} = {}& 1  \;,  \\
        \label{S1n}
	    S_{1,n} = {}& \frac{c_1\,c_2 \cdots c_{n-1}}{(b_n-b_1)(b_n-b_2)\cdots(b_n-b_{n-1})}  \;,  \quad n \ge 2  \;.
    \end{align}	
    It is not too difficult to show, by substituting \eqref{InvSmn}--\eqref{S1n} into the definition of $\beta_{k,n}$ in \eqref{gn} that 
    \begin{align}
    \label{gn(t)}
	    g_n(t) = \sum_{k=1}^n \, \beta_{k,n} \, e^{b_k\;\!t}  \;, 
    \end{align}
    where $\beta_{1,1} = 1$, and
    \begin{align}
    \label{Beta}
	    \beta_{k,n} = \left. c_1 \, c_2 \, \cdots \, c_{n-1} \middle/ \underset{j\ne k}{\prod_{j=1}^{n}}\,(b_k-b_j) \;, \quad n \ge 2 \right. \;.
    \end{align}
    Note the top limit of the sum in \eqref{gn(t)} is now $n$ rather than $K$ as a result of \eqref{InvSmn} for $m>n$.
    
    \subsection{Eigenvectors of $A$}
    \label{DerivationOfS}

    The goal here is to prove that $S$ has the form given in \eqref{Smn}. This is equivalent to showing that the $n$th column of $S$, which we denote by $\bm{v}^{(n)}$, satisfies
    \begin{align}
    \label{Av=bv}
	    A \, \bm{v}^{(n)} = b_n \, \bm{v}^{(n)}  \;,  \quad  n=1,2,\ldots,K  \;.
    \end{align}
    In component form, the left-hand side of \eqref{Av=bv} is
    \begin{align}
	    \big[ A \, \bm{v}^{(n)} \big]_m = {}& \sum_{j=1}^K \; \big( \, b_{m} \, \delta_{m,j} + c_m \, \delta_{m,j-1} \, \big) \, v^{(n)}_j  \\
                                  = {}& \sum_{j=1}^K \; b_{m} \, \delta_{m,j} \, S_{j,n} + \sum_{j=1}^K \; c_m \, \delta_{m,j-1} \, S_{j,n} 
                                  = b_{m} \, S_{m,n} + c_m \, S_{m+1,n}  
    \end{align}
    where we have noted that $v^{(n)}_j=S_{j,n}$ in the second equality. The eigenvalue equation \eqref{Av=bv}, when written in terms of the components of $S$ is therefore
    \begin{align}
    \label{ComponentAv=bv}
	    b_{m} \, S_{m,n} + c_m \, S_{m+1,n} = b_n \, S_{m,n}  \;.
    \end{align}
    We first show that $S$ satisfies \eqref{ComponentAv=bv} for all $m$ and $n$ such that $m<n$. Using \eqref{Smn} we have
    \begin{align}
    \label{Sm<n1}
	    b_{m} \, S_{m,n} + c_m \, S_{m+1,n} = b_m \; \frac{\prod_{k=m}^{n-1} c_k}{\prod_{k=m}^{n-1} \, (b_n-b_k)} 
	                                      + c_m \; \frac{\prod_{k=m+1}^{n-1}c_k}{\prod_{k=m+1}^{n-1} \, (b_n-b_k)}  \;.
    \end{align}
    Now we note that 
    \begin{align}
	    c_m \; \prod_{k=m+1}^{n-1} c_k = \prod_{k=m}^{n-1} c_k  \;,  \quad  \prod_{k=m+1}^{n-1} \, (b_n-b_k) = \frac{1}{b_n-b_m} \; \prod_{k=m}^{n-1} \, (b_n-b_k)  \;.
    \end{align}
    Equation \eqref{Sm<n1} thus becomes
    \begin{align}
    \label{Sm<n2}
	    b_{m} \, S_{m,n} + c_m \, S_{m+1,n} = {}& b_m \; \frac{\prod_{k=m}^{n-1} c_k}{\prod_{k=m}^{n-1} \, (b_n-b_k)} 
	                                          + (b_n-b_m) \; \frac{\prod_{k=m}^{n-1}c_k}{\prod_{k=m}^{n-1} \, (b_n-b_k)}  \\
                                      = {}& b_n \; \frac{\prod_{k=m}^{n-1}c_k}{\prod_{k=m}^{n-1} \, (b_n-b_k)} 	\\
                                      = {}& b_n \, S_{m,n}  \;.                                      
    \end{align}
    For $m>n$ it is clear that \eqref{Smn} gives zero for the left-hand of \eqref{ComponentAv=bv} and also zero for the right-hand side. Finally, for $m=n$ it is also trivial to see that \eqref{Smn} gives $b_m$ for both sides of \eqref{ComponentAv=bv}.
 
\section{Validity of the inverse-number expansion}
\label{Validity}

    \subsection{An iterative derivation of the inverse-number expansion and}

    To show that \eqref{E[Ups(t)]KthOrder}, \eqref{gn(t)MainText}, and \eqref{BetaMainText} of Sec.~\ref{InvNumExp} are indeed correct we consider an alternative way to derive the inverse-number expansion. Let us return to \eqref{dE[Ups(t)]/dt} and note that it could have been solved iteratively. That is, it has the formal solution
    \begin{align}
    \label{Recursion}
	    {\rm E}\big[\Upsilon^n(t)\big] = {\rm E}\big[\Upsilon^n(0)\big] \, e^{b_n\;\!t} + c_n \int^t_0 \, dt' \; {\rm E}\big[\Upsilon^{n+1}(t')\big] \, e^{b_n(t-t')}  \;.
    \end{align}
    This now gives us a recursive relation between any two consecutive moments of $\Upsilon(t)$ so we may write ${\rm E}[\Upsilon(t)]$ up to any desired order in moments of $\Upsilon(0)$ by repeated application of \eqref{Recursion}. For example, application of \eqref{Recursion} for $n=1,2,3$ gives
    \begin{align}
    \label{E[Ups(t)]Check}
	    {\rm E}\big[\Upsilon(t)\big] = {\rm E}\big[\Upsilon(0)\big] \, e^{b_1\;\!t} 
	                                   & + c_1 \int^t_0 \, dt_1 \; e^{b_2\;\!t_1} \;  {\rm E}\big[\Upsilon^{2}(0)\big] \nonumber \\
	                                   & + c_1 \, c_2  \int^t_0 \, dt_1 \int^{t_1}_0 \; dt_2  \; e^{b_3\;\!t_2} \, e^{b_2(t_1-t_2)} \; {\rm E}\big[\Upsilon^{3}(0)\big] \\
	                                   & + c_1 \, c_2 \, c_3  \int^t_0 \, dt_1 \int^{t_1}_0 \; dt_2 \int^{t_2}_0 \; dt_3 \; e^{b_3\;\!(t_2-t_3)} \, e^{b_2(t_1-t_2)} \; 
	                                     {\rm E}\big[\Upsilon^{4}(t_3)\big] \nonumber.
    \end{align}
    Using the definition of the expansion coefficients in \eqref{E[Ups(t)]KthOrder}, we find on repeated application of \eqref{E[Ups(t)]Check} the general result
    \begin{align}	
    \label{Coeffn=1}
	    g_1(t) = {}& e^{b_1\;\!t}  \;,  \\
    \label{Coeffn>2}
	    g_n(t_0) = {}& c_1 \, c_2 \cdots c_{n-1} \;  e^{b_1\;\!t} \int^{t_0}_0 dt_1 \; e^{(b_2-b_1)t_1} \int^{t_1}_0 dt_2 \; e^{(b_3-b_2)t_2}  
	               \cdots \nonumber \\
	               & \cdots \int^{t_{n-2}}_0 dt_{n-1} \; e^{(b_{n}-b_{n-1})t_{n-1}}  \;, \quad n \ge 2  \;.
    \end{align}
    Equations \eqref{Coeffn=1} and \eqref{Coeffn>2} may also be seen as an alternative form for the time-dependent coefficients of the inverse-number expansion. Compared to \eqref{gn(t)MainText} and \eqref{BetaMainText} they are rather clumsy to use for large $K$ as \eqref{Coeffn>2} requires one to compute an $n$-fold integral for $g_{n-1}(t)$. However, the integrals only ever involve simple exponentials so they are not too difficult to evaluate for small $n$. We can at least show that they match the results in \eqref{g2MainText} and \eqref{g3MainText}. Considering $n=2$ and $n=3$,
    \begin{align}
	    g_2(t) = {}& c_1 \, e^{b_1\;\!t} \int^t_0 dt_1 \; e^{(b_2-b_1)\,t_1}   \nn \\
    \label{g2Iterate}     
	       = {}& \frac{c_1}{b_2-b_1} \;  \big( \,e^{b_2\,t} - e^{b_1\,t} \big)  \;,  \\[0.25cm]
	    g_3(t) = {}& c_1 \, c_2 \, e^{b_1\;\!t} \int^t_0 dt_1 \; e^{(b_2-b_1)t_1} \int^{t_1}_0 dt_2 \; e^{(b_3-b_2)t_2}   \nn \\
    \label{g3Iterate}
	       = {}& \frac{c_1\,c_2}{(b_3-b_2)(b_3-b_1)} \; \big( e^{b_3\,t} - e^{b_1\;\!t} \big) 
	             - \frac{c_1\,c_2}{(b_3-b_2)(b_2-b_1)} \; \big( e^{b_2\,t} - e^{b_1\;\!t} \big)  \;.
    \end{align}
    It is clear that \eqref{g2Iterate} is identical to \eqref{g2MainText}. To see that \eqref{g3Iterate} is consistent with \eqref{g3MainText} we only have to add the coefficients of $\exp(b_1\;\!t)$ in \eqref{g3Iterate}.
    
    \subsection{Number and phase as stochastic processes}

    \begin{figure}[t]
    \centerline{\includegraphics[width=0.85\textwidth]{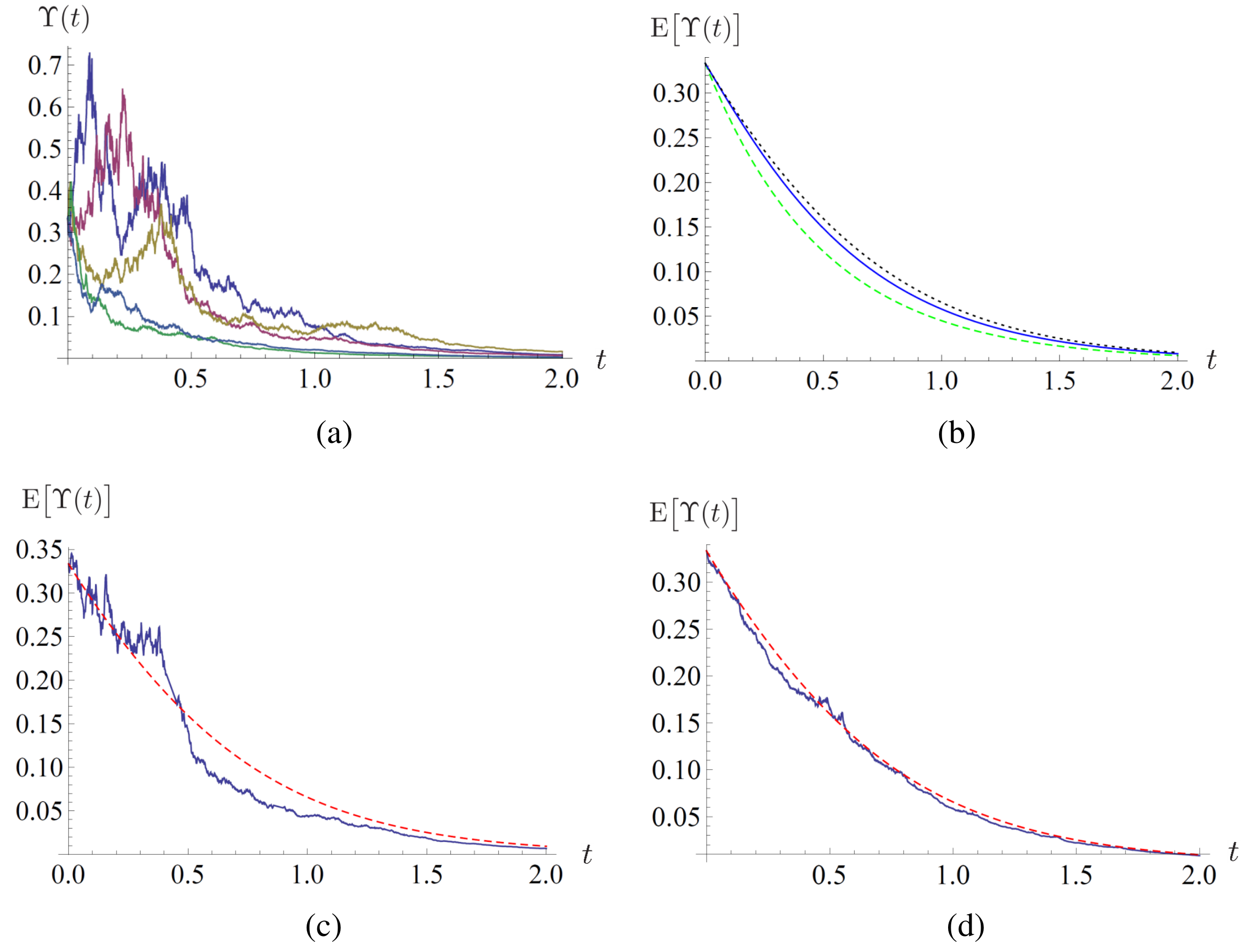}}
    \caption{\label{InverseNumberExpansion} Inverse number for a coherent-state input $\ket{\alpha}$ with $|\alpha|^2=\Upsilon(0)=3$ and $\kdn=0$ and $\kup=2$ (ideal linear amplifier). (a) Five samples of the process $\Upsilon(t)$ from simulating \eqref{dUpsMainText}. (b) Analytic expressions for the inverse-number expansion [\eqref{E[Ups(t)]KthOrder}, \eqref{gn(t)MainText} and \eqref{BetaMainText}] corresponding to $K=1$ (green dashed curve), $K=2$ (blue solid curve), and $K=3$ (black dotted curve). (c) Sample average for $\E[\Upsilon(t)]$ obtained from the five realisations in (a) (noisy blue curve) and the analytic formula corresponding to $K=3$ (red dashed curve). (d) $\E[\Upsilon(t)]$ obtained from averaging over 200 realisations of $\Upsilon(t)$ (solid blue curve), and from the inverse-number expansion for $K=3$ (red dashed curve).}
    \end{figure}
    We can illustrate how the inverse-number expansion works by comparing it to the average of $\Upsilon(t)$ over many realisations obtained from simulating \eqref{dUpsMainText}. This comparison is made in Fig.~\ref{InverseNumberExpansion} for the simple case of an ideal linear amplifier. We first show how different realisations of $\Upsilon(t)$ look like in Fig.~\ref{InverseNumberExpansion}(a). For visual clarity only five realisations of $\Upsilon(t)$ are shown. We then illustrate the analytic expression of $\E[\Upsilon(t)]$ for a few values of $K$. If the claimed time-dependent coefficients given in \eqref{gn(t)MainText} and \eqref{BetaMainText} are correct then we ought to find a value of $K$ for which there is a good agreement between the analytic expression for $\E[\Upsilon(t)]$ and that obtained from a large sample of $\Upsilon(t)$. In Fig.~\ref{InverseNumberExpansion}(b) we show how the sum in \eqref{E[Ups(t)]KthOrder} changes for $K=1$ (green dashed curve), $K=2$ (blue solid curve), and $K=3$ (black dotted curve). In Fig.~\ref{InverseNumberExpansion}(c) we compare the inverse-number expansion with $K=3$ with the average of the five realisations of $\Upsilon(t)$ shown in (a). The same comparison is made with 200 realisations in Fig.~\ref{InverseNumberExpansion}(d), and as can be seen, the sample average (solid blue line) is much smoother due to the large sample size and is in good agreement with the analytic result (red dashed line). We note that in practice there is some threshold value of $K$ beyond which the inverse-number expansion starts to deviate from the $\E[\Upsilon(t)]$ calculated by averaging over many sample paths though in theory we expect this $K$ to be arbitrarily large. We find that for a coherent-state input $\ket{\alpha}$, this threshold tend to be larger for larger $\alpha$. As long as $K$ is less than the threshold we find good agreement between the analytic result given by the inverse-number expansion and the stochastic simulations.
    
\section{Phase in quantum optics and the use of the $P$ distribution}
\label{PhaseWithinP}

    Here we comment on the nature of our treatment of phase and its relation to other work on phase diffusion in linear amplifiers. We have chosen to work with the stochastic differential equations that are consistent with the Glauber--Sudarshan function as opposed to other quasiprobability distributions. It has been shown explicitly for linear amplifiers that it is the \fp\ equation for the Wigner function (the $W$) rather than the Husimi (the $Q$), or the Glauber--Sudarshan (the $P$) which gives the correct phase diffusion coefficient in the small-noise approximation \cite{BSP89,Lu90}. This is perhaps not hard to accept given that $W(q,p)$, i.e.~when parametrised by the canonical position and momentum, guarantees true marginal distributions when either $q$ or $p$. The idea of using the Wigner function is that one would expect this to still be true if the Wigner function was reparametrised by number and phase (a more complete discussion of this point is given in Sec.~\ref{Conclusion} in connection to possible future directions). Then in what sense does the Glauber--Sudarshan distribution provide a valid description of the phase noise in linear amplifiers? If we use the $P$ rather than the $W$ then we forgo the transient dynamics of phase during amplification. The stationary statistics of phase however remain correct. It is already known that $P$, $W$, and $Q$ all give the same limiting phase distribution for an ideal linear amplifier in the high-gain regime \cite{Ban91,Pau74} (though there is no reason to expect this to breakdown for nonideal amplifiers). For a fixed amplification rate $\kup$, the high-gain regime simply means that we amplify the field for sufficiently long. With the assistance of Ref.~\cite{Ban91} we can actually compute the output phase probability density from the $P$ distribution explicitly for a coherent-state input $\ket{\alpha}$ in the high-gain limit. The result, for $\alpha=|\alpha|\exp(i\theta)$, is
    \begin{figure}[t]
        \centerline{\includegraphics[width=0.9\textwidth]{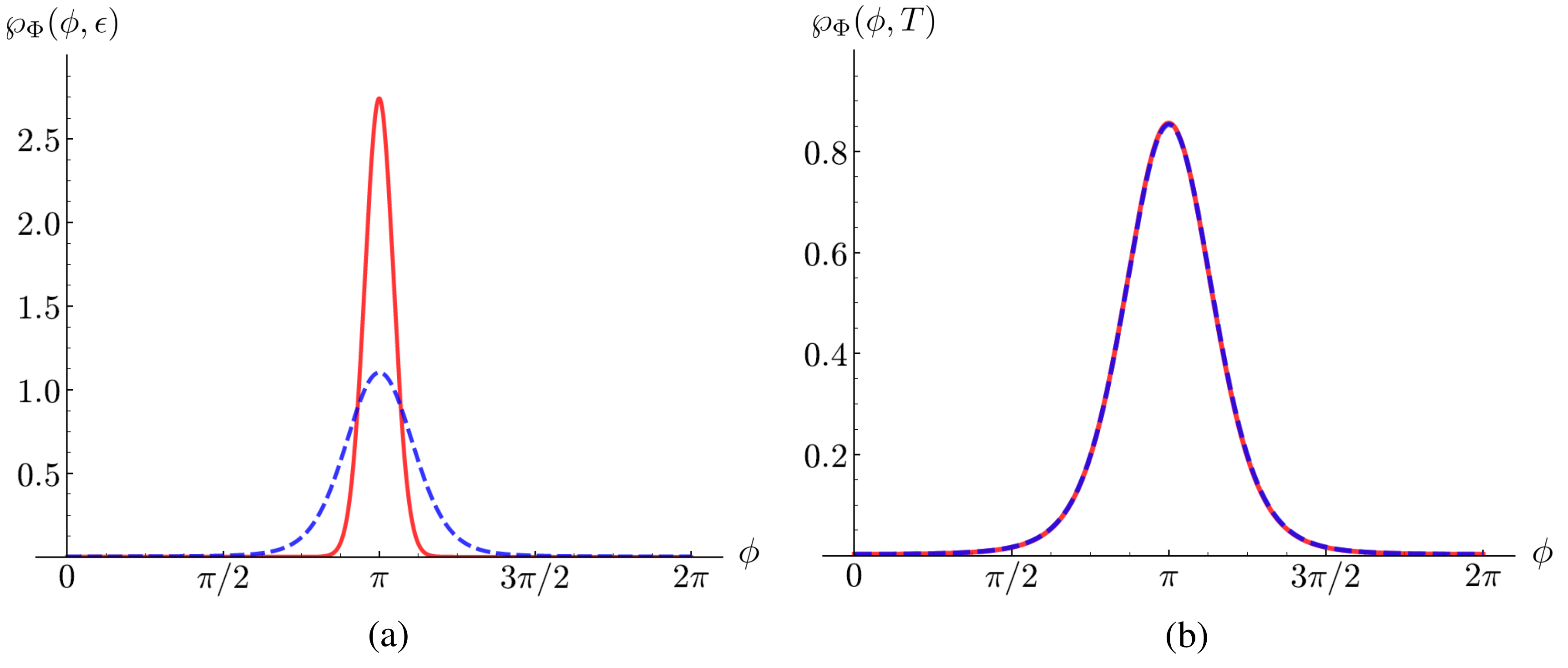}}
        \caption{\label{PhaseDistConvergence} Plots of $\wp_\Phi(\phi,t)$ as given by the Glauber--Sudarshan function (red solid line), and the Pegg--Barnett distribution (blue dashed line). The parameters are $\kup=1$, $\kdn=0$ and $\alpha=-1.5$. The two phase distributions are shown after (a) a short time, where $\epsilon=0.1$; and (b) after transience, where $T=4$.}
    \end{figure}
    \begin{align}
    \label{wpPhi(t)}
        \wp_\Phi(\phi,t) = {}& \frac{\cos\big(\phi-\theta\big)}{2\pi} \; e^{-\eta(t)}
                        \Big\{ \sqrt{\pi\,\eta(t)} \, \exp\!\big[ \,\eta(t) \, \cos^2\!\big(\phi-\theta\big) \big] \nonumber \\ 
                       & \times \left( 1 + \text{erf}\Big[ \sqrt{\eta(t)} \, \cos\big(\phi-\theta\big) \Big] \right) + \sec\!\big(\phi-\theta\big) \Big\} \;,
    \end{align}
    where ${\rm erf}(x)$ is the error function given by 
    \begin{align}
	    {\rm erf}(x) = \frac{2}{\rt{\pi}} \; \int^x_0 dy \; e^{-y^2}  \;,
    \end{align}
    and we have defined $\eta(t)$ to be the (noise-free) output signal normalised by the spontaneous-emission noise [see \eqref{E[N(t)]Exact} from Sec.~\ref{IntroMINexp}]
    \begin{align}
	    \eta(t) = \frac{G_t \, |\alpha|^2}{n_{\rm sp}(t)}  \;,  \quad  n_{\rm sp}(t) = G_t - 1   \;.
    \end{align}
    Note that here $G_t=\exp(\kup\,t)$ since an ideal linear amplifier is assumed. Using \eqref{wpPhi(t)} the phase statistics beyond transience can now be seen explicitly. To show that it converges to the correct phase distribution we use the Pegg--Barnett probability density as a reference. The Pegg--Barnett phase probability corresponding to some arbitrary state $\rho(t)$ is defined on a truncated $s$-dimensional Hilbert space spanned by the the Fock basis $\{\ket{n}\}_{n=0}^s$ \cite{BV07}:
    \begin{align}
	    \wp_{\Phi}(\phi_m,t) = \frac{s+1}{2\,\pi} \; \bra{\phi_m} \, \rho(t) \, \ket{\phi_m}  \;,
    \end{align}
    where for each $m$, the phase state $\ket{\phi_m}$ is defined by
    \begin{align}
	    \ket{\phi_m} = \frac{1}{\rt{s+1}} \; \sum_{n=0}^s \, e^{i n \phi_m} \, \ket{n} \;,  \quad  \phi_m = \phi_0 + \frac{2\pi}{s+1}\;m   \;.
    \end{align}
    \begin{figure}[t]
        \centerline{\includegraphics[width=0.5\textwidth]{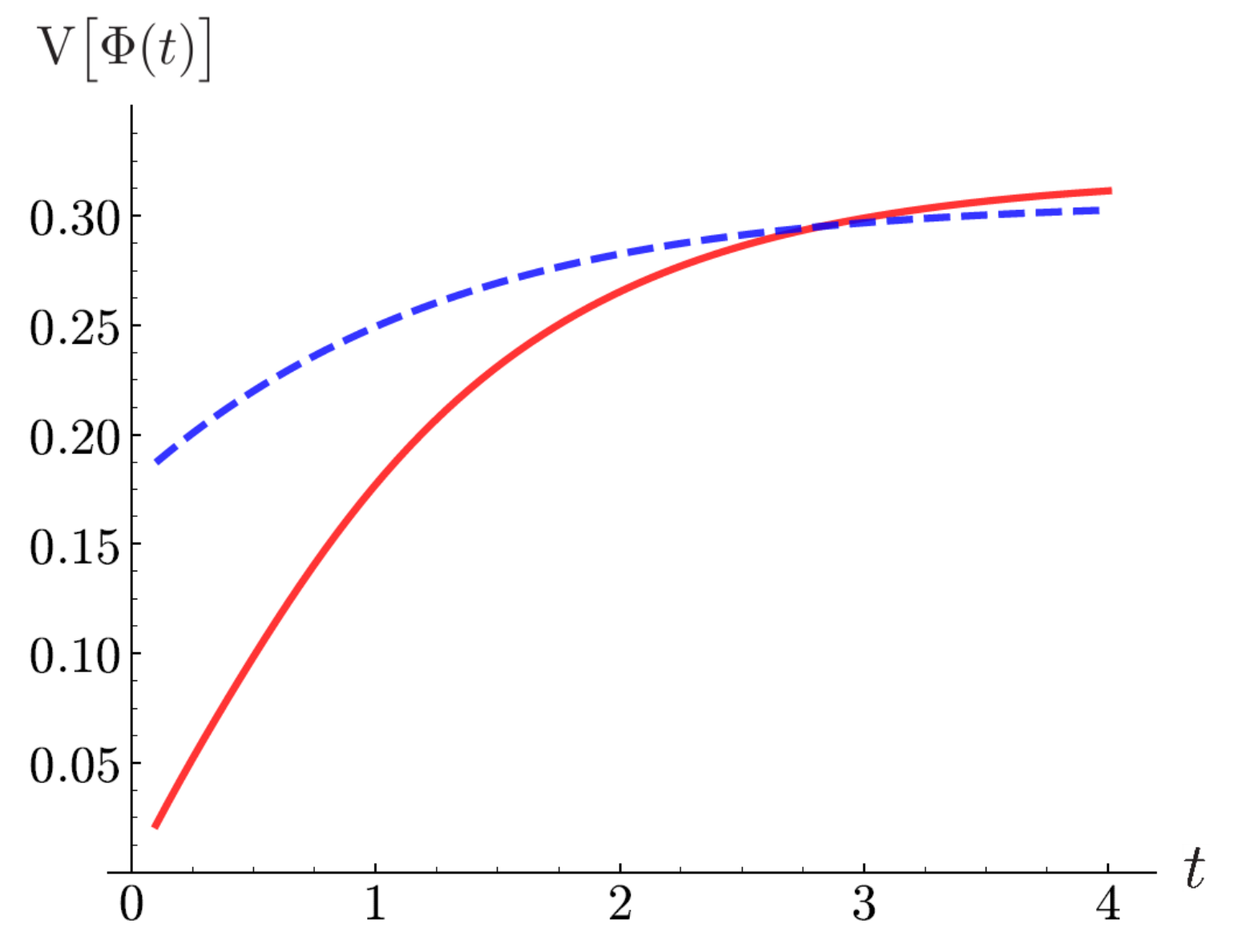}}
        \caption{\label{PhaseVarFromDist} Phase variance as a function of time as obtained from the $P$ function (red solid line) and the Pegg--Barnett distribution (blue dashed line). The parameter values are the same as those in Fig.~\ref{PhaseDistConvergence}.}
    \end{figure}
    Generally, all phase statistics obtained using the Pegg--Barnett operator must be calculated with a finite value of $s$, i.e.~on the truncated space first. To obtain the correct statistics we must then take the $s\longrightarrow\infty$ limit at the end (and only in the end). An exception to this limiting procedure applies for physical states which Pegg and Barnett defined to be states with finite average photon number. This is often true and is also true here. We obtained the Pegg--Barnett phase probability distribution numerically from the exact model given by the master equation \eqref{DensityOpEOM} in Sec.~\ref{Model}. We show the result of \eqref{wpPhi(t)} with the Pegg--Barnett distribution in Fig.~\ref{PhaseDistConvergence} at two times. In both Fig.~\ref{PhaseDistConvergence}(a) and (b) the Pegg--Barnett phase probability is shown as the blue dashed curve while \eqref{wpPhi(t)} is shown as the red solid curve. At $t=0$ the phase variance from the $P$ distribution would be zero. This is because the coherent state has a delta-function representation in terms of $P$. The Pegg--Barnett distribution on the other hand would correspond to the actual phase distribution which would have some finite width at $t=0$. After a short time $\epsilon$ the two distributions then broaden a little as shown in Fig.~\ref{PhaseDistConvergence}(a). The $P$-function phase density given by \eqref{wpPhi(t)} can be seen to be much more sharply peaked than the Pegg-Barnett phase density. This makes sense since it diffused from an initial delta function $\delta(\phi-\theta)$ at $t=0$. As time progresses we find the $P$-function phase density to diffuse further, eventually converging to the Pegg--Barnett distribution at time $T$ (the value of $T$ is taken to be four). We also show how the phase variance evolves in the interval $[\epsilon,T]$ in Fig.~\ref{PhaseVarFromDist}. It can be seen that the phase diffuses much more quickly when described in terms of the $P$ distribution (red solid curve). In relation to this we note that previous analysis have shown the Pegg--Barnett phase diffusion coefficient to be smaller than that from the $P$ function, but under the small-noise assumption \cite{BSP89}. In the end, it is clear that at $t=T=4$ the two phase variances differ only slightly, suggesting that the Glauber--Sudarshan distribution provides a good estimate for the phase variance after transience, or equivalently, in the high-gain regime (which is the limit in which one typically would like to operate the amplifier in \cite{CDG+10}). The $P$ distribution then has the advantage that coherent-state inputs are easily handled as it amounts to specifying deterministic initial conditions for the number and phase. In this case, all of the output noise is interpreted as added noise due to amplification as the input is completely noise-free.


\end{document}